\documentclass[aps,prx,reprint,superscriptaddress,amsmath,amssymb,floatfix,longbibliography]{revtex4-2}
\usepackage{graphicx}
\usepackage{mathtools}
\usepackage{amsthm}
\usepackage{bm}
\usepackage{xcolor}
\usepackage{hyperref}
\hypersetup{colorlinks=true,citecolor=blue,urlcolor=blue,linkcolor=blue,pdftitle={Gibbs-state phases beyond microscopic symmetry breaking: A phase invariant for the 3D toric code},pdfauthor={Haruki Watanabe}}
\newcommand{\Tr}{\operatorname{Tr}}
\newcommand{\id}{\mathbf 1}

\newtheorem{proposition}{Proposition}
\begin{document}

\title{Gibbs-state phases beyond microscopic symmetry breaking:\texorpdfstring{\\}{ }A phase invariant for the 3D toric code}
\author{Haruki Watanabe}
\email{hwatanabe@ust.hk}
\affiliation{Department of Physics, Hong Kong University of Science and Technology, Clear Water Bay, Hong Kong, China}
\affiliation{Institute for Advanced Study, Hong Kong University of Science and Technology, Clear Water Bay, Hong Kong, China}
\affiliation{Center for Theoretical Condensed Matter Physics, Hong Kong University of Science and Technology, Clear Water Bay, Hong Kong, China}
\date{\today}

\begin{abstract}
Do phases of Gibbs states admit a classification beyond symmetry breaking? We study the three-dimensional toric code with transverse and longitudinal fields that explicitly break both microscopic higher-form symmetries. We introduce the connected correlation $f_W$ of two locally decoded winding Wilson loops. Nonzero asymptotic $f_W$ rules out shallow local preparation from product states. An asymptotic optimization over the observables and physical recovery channels defines an invariant $\mathcal{F}$, which we prove constant within phases defined by two-way local channels with product ancillas and sublinear causal radius. The invariant equals one for the canonical zero-temperature state of the field-free toric code. Quantum Monte Carlo tests the conjectured thermodynamic values $f_W=1$ at weak fields and low temperature and $f_W=0$ elsewhere. The fixed-decoder correlation provides a lower bound on $\mathcal{F}$, so the results suggest an extended region with $\mathcal{F}=1$. Product states, including the infinite-temperature state and fully polarized infinite-field limits, have $\mathcal{F}=0$ exactly and therefore lie in a different phase from states with $\mathcal{F}=1$.
\end{abstract}
\maketitle

\section{Introduction}\label{sec:introduction}
Symmetry breaking provides a central organizing principle for phases of matter~\cite{landau1937theory,anderson1984basic,stanley1971introduction}. At zero temperature, intrinsic topological order~\cite{wen1990topological,kitaev2003fault,chen2010local}, quantized Hall responses~\cite{thouless1982quantized,haldane1988model}, and symmetry-protected topological phases~\cite{affleck1987rigorous,kennedy1992hidden,chen2012symmetry} demonstrate distinctions not captured by conventional symmetry-breaking order parameters. Extending this perspective to interacting Gibbs states at nonzero temperature motivates the search for phase distinctions that can be related both to local state preparation and to computable observables.

At finite temperature, the two-dimensional XY model already illustrates a transition without conventional symmetry breaking: its Berezinskii--Kosterlitz--Thouless transition separates regimes with algebraic and exponential correlations~\cite{berezinskii1971destruction,kosterlitz1973ordering,kosterlitz1974critical}. The absence of magnetic long-range order and the behavior of correlations have been established analytically~\cite{mermin1966absence,mcbryanSpencer1977,frohlichSpencer1981} and studied with high-precision Monte Carlo~\cite{hasenbusch2005}. The field-free three-dimensional toric code provides another established example: topological entropy distinguishes its low- and high-temperature Gibbs states~\cite{castelnovo2008topological}.

Here we study the three-dimensional toric code with transverse and longitudinal fields that together explicitly break both microscopic higher-form symmetries of the field-free model. By connecting a rigorous invariant under specified two-way local channels to a decoded Wilson-loop correlation accessible to quantum Monte Carlo, we obtain a practical witness for phase distinctions beyond microscopic symmetry breaking.

\begin{figure}[!tb]
\centering
\includegraphics[width=\columnwidth]{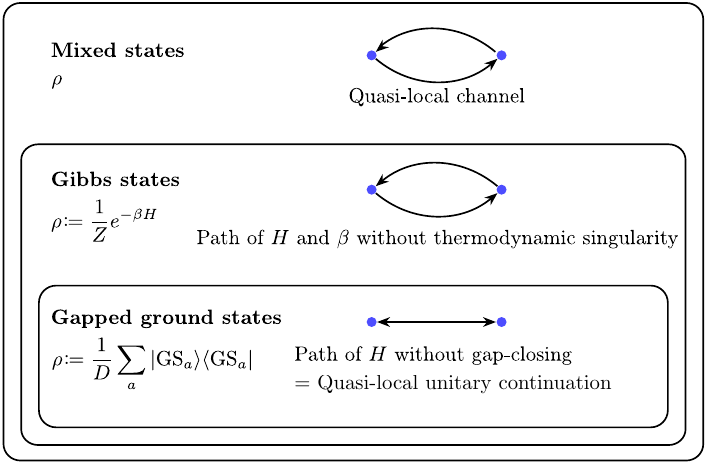}
\caption{\label{fig:tiers}
Classification questions for three state families. Gapped ground states motivate local-unitary equivalence; general mixed states admit comparisons through two-way local channels. The intermediate Gibbs family motivates paths without thermodynamic singularities. Here $D$ is the ground-state degeneracy, and $\{|\mathrm{GS}_a\rangle\}_{a=1}^D$ is an orthonormal basis of the ground-state subspace.}
\end{figure}

Figure~\ref{fig:tiers} places Gibbs states between gapped ground states and general mixed states. Ground-state phases admit local-unitary continuation, whereas general mixed states can be compared by local quantum channels~\cite{hastings2005quasiadiabatic,coser2019classification,sang2024mixed,rakovszky2024defining,xue2024tensor}. These channels also compare Gibbs states without requiring Gibbs intermediates. For a Hamiltonian $H$ at temperature $T$, the Gibbs state is $\rho\coloneqq e^{-\beta H}/Z$, where $\beta\coloneqq(k_BT)^{-1}$ is the inverse temperature, $k_B$ is the Boltzmann constant, and $Z\coloneqq\Tr e^{-\beta H}$ is the partition function. Within this Gibbs family, one can also ask whether temperature and the local Hamiltonian can be changed without a thermodynamic singularity. Smooth finite-volume connectivity is insufficient: finite systems always admit analytic Gibbs interpolations.

Local channels allow measurement and discarding, so on-site resets map every state to any product state. We therefore require two-way local-channel conversions~\cite{sang2024mixed}, with trace-norm errors vanishing as the linear size $L\to\infty$, using product ancillas and the controlled quasi-local extensions below. The channels need not be inverses. All product states share one trivial mixed-state phase, including the infinite-temperature state and fully polarized infinite-field limits at fixed finite temperature and field direction. Excluding this phase requires obstructing preparation from product inputs.

A bare winding Wilson loop is sensitive to small flux loops near its path, so local fluctuations can obscure the winding information it is intended to probe. We first apply a fixed local decoder that processes nearby flux defects before reading the winding parity. Its range grows with system size while remaining small compared with the separation of the two measured loops.

Our computable diagnostic, the decoded Wilson-loop correlation $f_W$, is the connected correlation of the two decoded Wilson-loop values. The connected subtraction distinguishes an unbiased shared classical winding bit from the agreement of two loop values pinned by polarization. Since the decoded operators are diagonal in the $Z$ basis, their full quantum Gibbs expectations can be evaluated from equal-time quantum Monte Carlo snapshots, including at nonzero transverse field. Using a single Sweep prescription, we test the conjectured thermodynamic value $f_W=1$ at weak fields and low temperature and $f_W=0$ elsewhere.

The first exact result is an obstruction to local preparation. Two distant decoded observables cannot acquire a connected correlation from initially independent spins if their preparation and decoding neighborhoods remain disjoint. We prove this statement for physical local circuits with product ancillas, including vanishing trace-norm preparation errors, and extend it to controlled quasi-local evolution. Nonzero asymptotic $f_W$ therefore excludes the trivial mixed-state phase. A zero of a fixed diagnostic does not establish local preparability, and its value can change within a single phase. An explicit on-site rotation illustrates this dependence and shows how to adapt the observables and recovery.

The second exact result is a phase invariant. Optimizing over observables in separated winding tubes and physical recovery channels, with a common radius bound and a specified order of limits, defines $\mathcal{F}$. We prove that it cannot increase under the allowed local channels and is therefore constant under two-way channel equivalence. It equals one for the canonical zero-temperature state of the field-free toric code and zero for all product-state families. The fixed-Sweep correlation supplies a lower bound: $f_W\to1$ implies $\mathcal{F}=1$. Thus the numerical results suggest a finite-temperature, weak-field region with the unit value, whereas suppressed $f_W$ elsewhere is only consistent with a zero optimized value. The invariant can also detect conventional long-range correlations; it is neither a complete phase classification nor a diagnostic specific to intrinsic topological order.

Our use of local recovery connects to several approaches to mixed-state order. These use conditional mutual information~\cite{sang2025stability,zhang2025conditional}, locally reversible channels~\cite{sang2025reversibility}, and axioms of local recoverability~\cite{yangShiLee2025topological}. Decoherence-induced symmetry breaking has been studied through nonlinear density-matrix correlations~\cite{lee2023quantum} and through strong-to-weak symmetry breaking, in which a symmetry acting separately on the ket and bra is reduced to a symmetry of their joint action~\cite{lessa2025strong,zhang2025swssb}. Connections between mixed-state order and error recovery have also been established for decohered topological codes~\cite{bao2023mixed,fan2024diagnostics,jongyeonLee2024coherent} and for fault-tolerance transitions characterized by spacetime Markov length~\cite{negari2024spacetime}.

The relation between channel equivalence and thermodynamic regularity is a further question. Two-way connectivity is expected within broad classes of noncritical thermal phases; approximate channel circuits can be constructed when stable clustering holds along a connecting Gibbs-state path within a fixed local interaction algebra~\cite{ma2025circuit}. This is a correlation-decay condition robust to the allowed local perturbations. Deriving this condition from free-energy analyticity in general remains open. Quantum belief propagation~\cite{hastings2007quantum} and recovery maps for small shielded regions~\cite{chenRouze2025} provide related tools for studying Gibbs states; these local statements alone do not supply the preparation bounds needed here for growing winding-loop supports.

Section~\ref{sec:model} introduces the model and its closed-flux constraint. Section~\ref{sec:decoded} defines the decoded Wilson-loop correlation and explains the decoder step by step. Section~\ref{sec:preparation} proves the obstruction to local preparation, and Sec.~\ref{sec:fw-numerics} presents the quantum Monte Carlo results. Section~\ref{sec:readout-transformations} explains the dependence on a fixed decoder and how to adapt it along a change of basis. Section~\ref{sec:phase-invariant} constructs the optimized phase invariant, proves its invariance, and relates its exact reference values to the numerical evidence. Section~\ref{sec:discussion} discusses the resulting phase and memory interpretations.

\section{Model and symmetries}\label{sec:model}

This section introduces the model, its established limiting transitions, and the closed-flux constraint that survives explicit symmetry breaking and supplies the input to decoding.

\subsection{Hamiltonian and established limits}
Place one qubit on each link $\ell$ of a periodic cubic lattice of linear size $L$. We suppress system-size dependence in the notation; unless otherwise specified, asymptotic statements refer to $L\to\infty$. The Hamiltonian is
\begin{equation}
H\coloneqq-J_e\sum_v A_v-J_m\sum_p B_p-h_x\sum_\ell X_\ell-h_z\sum_\ell Z_\ell.
\label{eq:H}
\end{equation}
Here $v$ and $p$ label vertices and elementary plaquettes, respectively; $\partial p$ consists of the four links bounding $p$. The star and plaquette operators are
\begin{equation}
A_v\coloneqq\prod_{\ell\ni v}X_\ell,\qquad
B_p\coloneqq\prod_{\ell\in\partial p}Z_\ell,
\label{eq:stabilizers}
\end{equation}
with $X_\ell,Z_\ell$ the Pauli operators on link $\ell$. Their coupling strengths are $J_e$ and $J_m$. We set $J_e=J_m=k_B=1$ and take nonnegative fields. The transverse term creates magnetic-flux loops by flipping four neighboring plaquettes; the longitudinal term creates electric charges by flipping two adjacent stars. At $h_x=h_z=0$, closed flux loops proliferate at $T_c^{(0)}\simeq1.3133$ through a second-order thermal transition with three-dimensional Ising criticality~\cite{wegner1971duality,nussinov2008thermal,castelnovo2008topological,pelissetto2002critical}.

Castelnovo and Chamon distinguished these field-free thermal phases using topological entanglement entropy (TEE) constructed from von Neumann entropies. In their normalization, this quantity, denoted by $S_{\rm topo}^{\rm 3D}(T)$, has the thermodynamic-limit form~\cite{castelnovo2008topological}
\begin{equation}
S_{\rm topo}^{\rm 3D}(T)=
\begin{cases}
2\ln 2, & T=0,\\
\ln 2, & 0<T<T_c^{(0)},\\
0, & T>T_c^{(0)}.
\end{cases}
\label{eq:field-free-tee}
\end{equation}
Thus only half of its zero-temperature value survives at arbitrarily small positive temperature.

At zero temperature, the toric-code gap is stable against sufficiently weak local perturbations~\cite{bravyi2010topological}. At $T=h_z=0$, a first-order quantum transition occurs at $h_x=1$. At $T=h_x=0$, a second-order quantum transition occurs at $h_z\simeq0.194$, with four-dimensional Ising$^*$ criticality~\cite{reiss2019quantum}. The asterisk distinguishes the gauge-theory operator content from that of the dual Ising model.

The thermal transition of the field-free model also places this problem in the broader study of thermal stability in quantum spin liquids~\cite{savaryBalents2017}. Quantum Monte Carlo has been used to study three-dimensional toric codes with ferromagnetic Ising interactions~\cite{kamiya2015magnetic}, and finite-temperature transitions have been studied in three-dimensional Kitaev spin liquids on hyperhoneycomb and hyperoctagon lattices~\cite{nasu2014finite,nasu2014vaporization,mishchenko2017finite}. Here we use the two fields in Eq.~\eqref{eq:H} to examine what remains when the microscopic higher-form symmetries are explicitly broken.

\subsection{Microscopic symmetries}

The relevant microscopic symmetries have generators
\begin{equation}
U_e(\Sigma^*)\coloneqq\prod_{\ell\perp\Sigma^*}X_\ell
\label{eq:electric-symmetry}
\end{equation}
for closed dual surfaces $\Sigma^*$ and
\begin{equation}
U_m(C)\coloneqq\prod_{\ell\in C}Z_\ell
\label{eq:magnetic-symmetry}
\end{equation}
for closed primal loops $C$. They generate the electric $\mathbb Z_2$ one-form and magnetic $\mathbb Z_2$ two-form symmetries~\cite{gaiotto2015generalized,mcgreevy2022generalized,lake2018higher}. The longitudinal field breaks the former and the transverse field breaks the latter. With both fields nonzero, neither microscopic higher-form symmetry remains. Lattice symmetries remain, but their breaking is not the distinction sought here.

This choice of perturbations separates our question from symmetry-based criteria for topological order~\cite{nussinov2009symmetry} and from thermal phases protected by higher-form symmetries~\cite{roberts2017symmetry}. Explicit breaking at the microscopic level does not, however, preclude an emergent higher-form description at long distances~\cite{paceWen2023exact}.

The gauge-theory context also separates deconfinement from conventional symmetry breaking. Local gauge symmetry cannot spontaneously break without gauge fixing~\cite{elitzur1975impossibility}, and analyticity results allow confinement and Higgs regimes to connect in gauge theories with fundamental matter~\cite{osterwalder1978gauge,fradkin1979phase}. The phase structure and multicriticality of the three-dimensional classical $\mathbb Z_2$ gauge--Higgs model, also relevant to the $(2+1)$-dimensional quantum toric code, have been studied extensively~\cite{tupitsyn2010topological,somoza2021self,bonati2022multicritical}. These results provide context rather than a phase diagram for the present three-dimensional quantum model.

\subsection{Closed-flux constraint}

Although the fields break the microscopic symmetries, they leave the flux constraint intact. The Bianchi identity
\begin{equation}
\prod_{p\subset\partial c}B_p=\id
\label{eq:bianchi}
\end{equation}
holds on each cube $c$, where $\id$ is the identity operator. Plaquette defects therefore have no flux endpoints: the dual edges piercing violated plaquettes form closed flux loops. Physical spin configurations on the torus also require an even number of flux-line crossings through a full cross-section perpendicular to each periodic direction. Individual loops may wind around the torus, but their winding parities must cancel in each direction. This is a kinematic constraint, not in itself a phase distinction.

The distinction between a constraint and an energetic suppression of defects matters here. Related spin-ice model studies show how thermally excited monopoles screen an emergent gauge field and round continuous transitions of the monopole-free limit into crossovers~\cite{watanabe2026ice,watanabe2026pyrochlore,watanabe2026arbitraryspin}. Here the absence of flux endpoints is an exact operator constraint at every temperature and field strength. The decoder defined next uses this closed-flux structure to extract winding information from individual spin configurations.

\section{Decoded Wilson-loop correlation}\label{sec:decoded}

In this section, we define $f_W$ and its snapshot estimator, explain the connected subtraction, and specify the Sweep decoder and its locality.

\subsection{Loop geometry and recovery channel}
\begin{figure*}[!t]
\centering
\includegraphics[width=\textwidth]{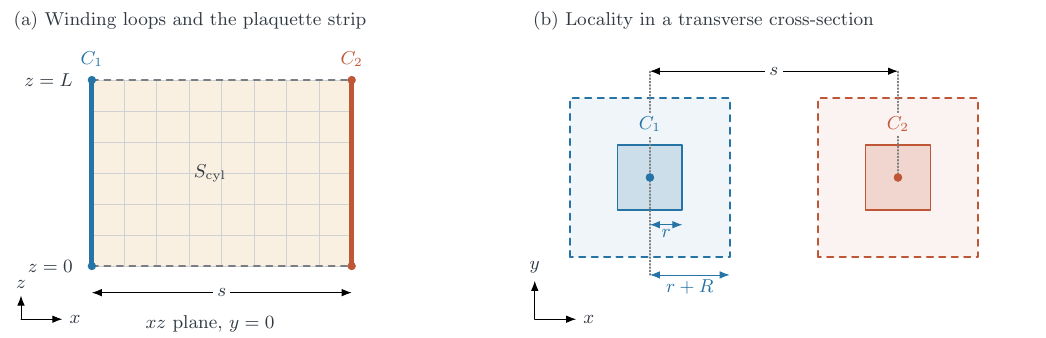}
\caption{\label{fig:loop-geometry}
Geometry and locality of the decoded Wilson-loop observables. (a) The two primal loops $C_1,C_2$, carrying $W_1,W_2$, bound the strip $S_{\rm cyl}$ of $xz$ plaquettes. Identifying $z=0$ and $z=L$ closes each loop and makes the strip cylindrical; the dashed horizontal edges are identified, not additional boundaries. (b) Viewed along $z$, the loops appear as points. Solid squares bound the radius-$r$ supports of the decoded observables. Dashed squares bound the radius-$(r+R)$ supports obtained by pulling them back through a preparation circuit of causal radius $R$. Their separation gives the product-input factorization in Sec.~\ref{sec:preparation}. The drawings are schematic; squares reflect the $L^\infty$ distance convention.}
\end{figure*}

Choose two parallel $z$-winding loops $C_1,C_2$, based at $(0,0,0)$ and $(\lfloor L/2\rfloor,0,0)$, with separation $s\coloneqq\lfloor L/2\rfloor$, where $\lfloor x\rfloor$ denotes the greatest integer not exceeding $x$. Figure~\ref{fig:loop-geometry}(a) illustrates this geometry. Define the bare Wilson operators along these loops by
\begin{equation}
W_i\coloneqq\prod_{\ell\in C_i}Z_\ell.
\label{eq:bare-wilson}
\end{equation}
Let $S_{\rm cyl}$ be the rectangular strip of $xz$ plaquettes joining the loops. Periodicity in $z$ makes it cylindrical, with boundary components $C_1$ and $C_2$. The identity
\begin{equation}
W_1W_2=\prod_{p\in S_{\rm cyl}}B_p
\label{eq:bare-pair}
\end{equation}
shows why small flux loops near the boundary can change the bare pair even when long-distance winding information survives. We use decoding to suppress these local contributions before measuring the winding information~\cite{dennis2002topological,cong2024enhancing}.

Let $\eta$ denote the joint outcome of measuring the plaquette operators, with $\eta_p=\pm1$. A deterministic decoder selects a set of links $\mathcal R(\eta)$ to flip. Syndrome measurement followed by this conditional Pauli correction defines the recovery channel
\begin{equation}
\mathcal C(\rho)\coloneqq\sum_\eta X_{\mathcal R(\eta)}P_\eta\rho P_\eta X_{\mathcal R(\eta)},
\label{eq:recovery}
\end{equation}
where
\begin{equation}
P_\eta\coloneqq\prod_p\frac{\id+\eta_pB_p}{2},\qquad
X_{\mathcal R}\coloneqq\prod_{\ell\in\mathcal R}X_\ell.
\label{eq:recovery-components}
\end{equation}
Completeness of the syndrome projectors makes this a completely positive, trace-preserving map. The recovery prescription is fixed before comparing states. For any state $\sigma$, write $\langle O\rangle_\sigma\coloneqq\Tr(\sigma O)$. We define the decoded Wilson-loop correlation as the connected correlation of the two loops in the recovered state:
\begin{equation}
f_W\coloneqq\langle W_1W_2\rangle_{\mathcal C(\rho)}
-\langle W_1\rangle_{\mathcal C(\rho)}\langle W_2\rangle_{\mathcal C(\rho)}.
\label{eq:fw}
\end{equation}
We write $f_W(\rho)$ when the state needs to be explicit.

\subsection{Decoded operators and their snapshot values}

The Wilson-loop statistics after recovery can equivalently be described by observables acting on the input state. The Heisenberg adjoint $\mathcal C^\dagger$ is defined by
\begin{equation}
\Tr[\mathcal C(\rho)O]=\Tr[\rho\,\mathcal C^\dagger(O)]
\label{eq:adjoint}
\end{equation}
for every state $\rho$ and observable $O$: it transfers the action of the channel from the state to the observable. Define the decoded operators by
\begin{equation}
\begin{aligned}
\widetilde W_i\coloneqq\mathcal C^\dagger(W_i)
&=\sum_\eta P_\eta X_{\mathcal R(\eta)}W_iX_{\mathcal R(\eta)}P_\eta\\
&=W_i\sum_\eta(-1)^{\mathcal R(\eta)\cdot C_i}P_\eta.
\end{aligned}
\label{eq:decoded-operators}
\end{equation}
The dot counts the links shared by $\mathcal R(\eta)$ and $C_i$. Because $[P_\eta,W_i]=0$ and $P_\eta P_{\eta'}=\delta_{\eta\eta'}P_\eta$, multiplication gives
\begin{equation}
\mathcal C^\dagger(W_1)\mathcal C^\dagger(W_2)
=\mathcal C^\dagger(W_1W_2).
\label{eq:decoded-product}
\end{equation} This multiplicativity is specific to the deterministic Pauli recovery; it need not hold for an arbitrary channel. Define $\widetilde W_{12}\coloneqq\mathcal C^\dagger(W_1W_2)$. The sign factors square to one and $\sum_\eta P_\eta=\id$, so also $\widetilde W_i^2=\id$. In the input state, Eq.~\eqref{eq:fw} becomes
\begin{equation}
f_W=\langle\widetilde W_1\widetilde W_2\rangle_\rho
-\langle\widetilde W_1\rangle_\rho\langle\widetilde W_2\rangle_\rho
=D_{12}-M_1M_2,
\label{eq:fw-input}
\end{equation}
where $D_{12}\coloneqq\langle\widetilde W_1\widetilde W_2\rangle_\rho$ and $M_i\coloneqq\langle\widetilde W_i\rangle_\rho$.

The decoded operators are diagonal in the $Z$ basis. For a complete $Z$-basis snapshot $z\coloneqq\{z_\ell=\pm1\}$ of all link spins, let $|z\rangle$ be the simultaneous eigenstate satisfying $Z_\ell|z\rangle=z_\ell|z\rangle$. The corresponding plaquette and Wilson-loop eigenvalues are
\begin{equation}
B_p(z)\coloneqq\prod_{\ell\in\partial p}z_\ell,\qquad
W_i(z)\coloneqq\prod_{\ell\in C_i}z_\ell.
\label{eq:snapshot-eigenvalues}
\end{equation}
The syndrome is $\eta_p(z)\coloneqq B_p(z)$, and the decoded operators have the binary snapshot values
\begin{equation}
w_i(z)\coloneqq\langle z|\widetilde W_i|z\rangle
=W_i(z)(-1)^{\mathcal R(\eta(z))\cdot C_i}\in\{\pm1\}.
\label{eq:readouts}
\end{equation}
Here $w_i(z)$ is the decoded Wilson-loop value for an individual snapshot, whereas $\widetilde W_i$ is the corresponding observable. The diagnostic $f_W$ is the connected correlation of $w_1,w_2$ over the snapshot distribution, not an individual measurement result. Both loop values use the same syndrome-dependent correction. Using the measurement probabilities $\langle z|\rho|z\rangle$, the moments are
\begin{equation}
\begin{aligned}
D_{12}&=\sum_z\langle z|\rho|z\rangle w_1(z)w_2(z),\\
M_i&=\sum_z\langle z|\rho|z\rangle w_i(z).
\end{aligned}
\label{eq:snapshot-moments}
\end{equation}
Thus the input-state covariance can be evaluated directly from complete spin snapshots.

\subsection{Flux interpretation and the role of the connected subtraction}

Decoding has a direct flux interpretation. For a snapshot $z$, call the dual edge piercing a plaquette $p$ occupied when $B_p(z)=-1$. This edge joins the centers of the two primal cubes sharing $p$. The defect graph consists of all occupied dual edges and their endpoints: its vertices are these cube centers, with vertices incident to no occupied edge omitted. Write $E$ for its edge set; the vertex set is then implicit. The Bianchi identity makes the degree, or number of occupied incident edges, of every such vertex $2$, $4$, or $6$. A connected component is a maximal part whose vertices are joined by paths of occupied edges. If every vertex of a component has degree two, it is a simple closed flux loop. At degree-four or degree-six vertices, a decomposition into individual loops need not be unique; the defect graph itself remains unambiguous.

Regard the correction links $\mathcal R$ as the dual faces they pierce, and let $\partial\mathcal R$ consist of the dual edges incident to an odd number of these faces. The residual defect graph has edge set $E'\coloneqq E\mathbin\triangle\partial\mathcal R$, where $\triangle$ is symmetric difference. After recovery, the product of plaquette values over $S_{\rm cyl}$ reduces to $w_1w_2$ because each interior link contributes twice. Thus
\begin{equation}
w_1w_2=(-1)^{|E'\cap S_{\rm cyl}|},\qquad
D_{12}=1-2P_{\rm fail},
\label{eq:residual}
\end{equation}
where $|E'\cap S_{\rm cyl}|$ counts residual flux lines piercing the strip and $P_{\rm fail}\coloneqq\Pr(w_1\ne w_2)$. A recovery that removes every defect would force $D_{12}=1$ even if it needed information from the entire lattice. Restricting the recovery to local operations is therefore essential to interpreting the pair as locally recoverable information.

The connected subtraction is essential. When the decoded loop values $w_1,w_2$ agree, they can represent a shared classical winding bit; unbiased agreement gives $D_{12}=1$, $M_i=0$, and $f_W=1$. Two loop values pinned to the same sign also give $D_{12}=1$, but now $M_1M_2=1$ and $f_W=0$. For example, the ferromagnetic product state with every $Z_\ell=+1$ has no plaquette defects, so the Sweep decoder below applies no correction and $w_1=w_2=+1$. The diagnostic thus distinguishes recoverable correlations from fixed winding parities due to polarization.

A complementary exact example is the normalized projector onto the field-free ground-state subspace at finite $L$. Every plaquette has $B_p=+1$, so the Sweep decoder applies no flips and $W_1W_2=1$. The equally weighted logical winding sectors give $M_1=M_2=0$, and hence $f_W=1$. This reference state illustrates the intended statistics of $w_1,w_2$; it does not by itself establish a nonzero-temperature limit uniform in $L$.

Since $w_i^2=1$, the centered random variable $w_i-M_i$ has variance $1-M_i^2$. Applying Cauchy--Schwarz to these two centered random variables gives the binary covariance bound
\begin{equation}
|f_W|\leq\sqrt{(1-M_1^2)(1-M_2^2)}\leq1.
\label{eq:covariance-bound}
\end{equation} This also gives the converse: $f_W\to1$ requires $M_i\to0$ and $D_{12}\to1$, namely unbiased loop values that agree with probability tending to one. At $h_z=0$, a logical membrane flips both loops without changing the syndrome, so $M_i=0$ in the full periodic Gibbs state. At nonzero $h_z$, both one-point functions must be measured rather than set to zero by symmetry.

The three moments determine the complete binary distribution:
\begin{equation}
\Pr(w_1=a,w_2=b)
=\frac{1+aM_1+bM_2+abD_{12}}{4},
\label{eq:joint}
\end{equation}
where $a,b=\pm1$. In particular, $f_W=0$ means independence of these two binary variables, not a product many-body state. The proposed $1/0$ limits concern the full Gibbs ensemble at fixed $(T,h_x,h_z)$ as $L\to\infty$, with the depth prescribed below and no conditioning on winding sectors. They do not assert quantization for arbitrary states. Section~\ref{sec:ferromagnetic-mixtures} shows that the optimized invariant is likewise not generally quantized.

\begin{figure*}[!t]
\centering
\includegraphics[width=\textwidth]{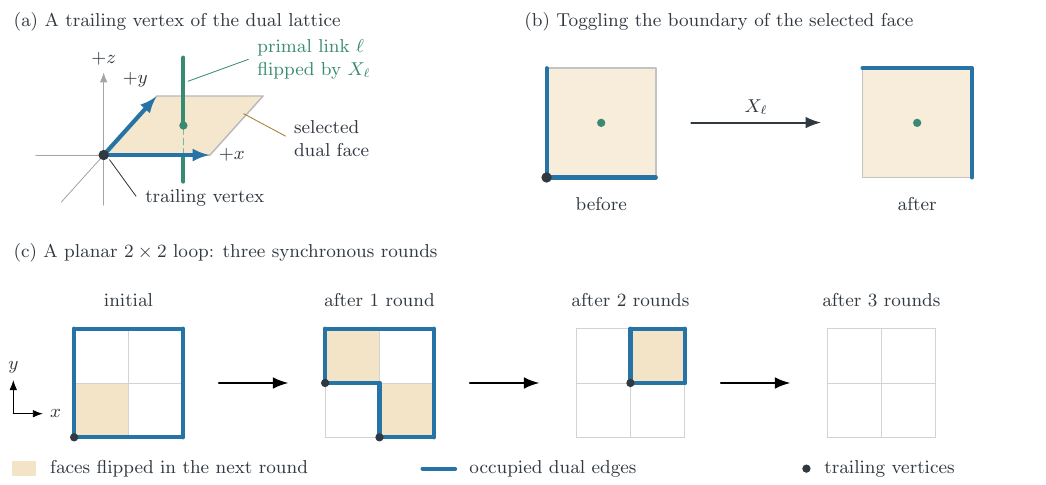}
\caption{\label{fig:sweep-rule}
The fixed local Sweep decoder. (a) The black dot is a trailing dual vertex: its occupied incident edges (blue) point along $+x$ and $+y$, with none in the negative directions. These two edges select the shaded dual face. Its piercing primal link $\ell$ (green) carries the spin flipped by $X_\ell$; all other depicted vertices, edges, and faces are dual. The dashed green segment lies behind the face. (b) This flip toggles the four boundary edges of the selected face. The rest of the defect graph is omitted. (c) An isolated planar $2\times2$ flux loop disappears in three synchronous rounds. At each stage, black dots identify trailing vertices and shading marks the faces selected for the next round. The two faces after the first round are flipped simultaneously using that stage's syndrome.}
\end{figure*}

\subsection{The fixed local Sweep decoder}\label{sec:sweep}
Figure~\ref{fig:sweep-rule} illustrates the local rule and its synchronous application. For the numerical calculations in this article, we choose the cubic-lattice Sweep decoder~\cite{kubica2019cellular,zhou2025finite} as one implementation of the recovery. Its updates act directly on the defect graph defined above and require no choice of a decomposition into loops. Orient the dual edges along $+x,+y,+z$, so each dual vertex has three outgoing edges in these positive directions and three incoming edges from the negative directions. This local convention also applies across periodic boundaries. A dual vertex is \emph{trailing} when at least one incident edge is occupied and none of its three edges in the negative directions is occupied. The Bianchi identity requires an even number of occupied edges at every dual vertex. Thus a trailing vertex has exactly two occupied outgoing edges, since only three outgoing edges are available.

These two edges select the unique elementary dual square that contains both of them. For example, the $+x$ and $+y$ edges select the dual $xy$ square on the positive sides of the vertex. This square is pierced by a $z$-directed link of the original lattice. Applying $X_\ell$ to the spin on that original link toggles the occupancy of all four dual edges on the square's boundary, removing the two occupied edges at the selected vertex. Each round evaluates all dual vertices using the same pre-round syndrome and applies the selected flips synchronously; vertices that are not trailing select no flip. Link flips combine modulo two across vertices and rounds to give $\mathcal R(\eta)$.
For a simple planar example, an isolated $1\times1$ defect loop disappears in one round. As shown in Fig.~\ref{fig:sweep-rule}(c), an isolated $2\times2$ square loop disappears in three synchronous rounds: one corner square is first flipped, then its two neighboring squares, and finally the remaining square. This illustrates why a round must evaluate all vertices from the configuration at its beginning.

\subsection{Decoder depth, locality, and erasure scale}
We use
\begin{equation}
r\coloneqq\min\left\{\lfloor\sqrt L\rfloor,
\left\lfloor\frac{s-1}{2}\right\rfloor\right\}
\label{eq:depth}
\end{equation}
rounds at every temperature and field. Since $s=\lfloor L/2\rfloor$, the second term grows as $L/4$, faster than $\sqrt L$. Thus $r=\lfloor\sqrt L\rfloor$ for all $L\geq18$; the second term only limits the depth at small system sizes, ensuring $2r<s$ there as well. For $L=6,8,10,12,14$, the minimum is given by the second term, yielding depths $1,1,2,2,3$, respectively.

\subsubsection{Support of the decoded observables}

To determine the information needed to compute $w_i(z)$, anchor each primal link at its starting vertex when oriented along a positive coordinate direction. Distances between anchors are measured with the periodic $L^\infty$ metric: take the shortest periodic distance in each coordinate, then the largest of the three. Each primal link pierces one dual square. Only the corner from which both sides of this square point in positive directions can select that link for a flip. The decision uses the six plaquettes surrounding this dual vertex, namely the six faces of one primal cube. Every spin entering those plaquettes has its anchor within distance one of the flipped link's anchor. One synchronous round therefore enlarges the dependence on the input spins by at most one cell. Iterating this statement, a spin after $r$ rounds depends only on input spins within distance $r$. Taking the product along $C_i$ shows that $w_i(z)$, and hence $\widetilde W_i$, is supported in a radius-$r$ tube around $C_i$.

The growing depth $r\to\infty$ allows larger local flux patterns to be processed, while $r/L\to0$ and $2r<s$ keep the support neighborhoods of the two decoded operators separate. Whole-lattice checks of flux-loop closure and the first identity in Eq.~\eqref{eq:residual} validate the output without influencing correction decisions.

\subsubsection{Erasure of isolated defect components}

A sufficient erasure criterion can be stated without decomposing the defect graph into loops. Let $n_K$ be the number of occupied edges in a nonempty connected component $K$ of the initial defect graph, counting each edge once even when loops touch. For a simple loop, $n_K$ is its length in lattice units. Assume that its coordinate bounding box has no periodic self-identifications and that other components do not affect its evolution before erasure. The number $N_K$ of synchronous rounds needed to erase it satisfies
\begin{equation}
N_K\leq\frac{n_K}{2}-1.
\label{eq:sweep-loop-length}
\end{equation}
To prove this bound, let $d_x,d_y,d_z$ be the initial coordinate spans of $K$, namely the differences between the largest and smallest vertex coordinates along each axis. Every selected dual face remains inside the initial bounding box. Moreover, all occupied vertices minimizing $x+y+z$ are trailing and disappear in the next round; no simultaneous update can reintroduce a vertex at that height. Thus the minimum occupied height increases by at least one per round. A nonempty even-degree defect graph must have two occupied outgoing edges at its minimum-height vertex, placing that vertex at least two height units below the upper corner of the box. Consequently, erasure takes at most $d_x+d_y+d_z-1$ rounds, even if the component splits during the updates. Because the initial $K$ is connected and every vertex has even degree, it admits a closed walk that traverses each edge exactly once. This walk crosses each coordinate span in both directions, giving $2(d_x+d_y+d_z)\leq n_K$ and proving Eq.~\eqref{eq:sweep-loop-length}.

Under these isolation and boundary assumptions, $r$ rounds therefore erase any component with $n_K\leq2(r+1)$. For a simple loop this is a loop-length criterion; for touching loops the combined edge count enters the bound. An axis-aligned planar $a\times b$ rectangular loop takes exactly $a+b-1$ rounds, saturating the bound because $n_K=2(a+b)$. Other shapes can disappear faster, so this criterion does not set the longest loop that can be erased. Applying it to a full Gibbs snapshot additionally requires control of interactions between components during decoding.

\subsection{Other syndrome-conditioned decoders}\label{sec:decoder-choices}

The construction allows other decoders. A sufficient class for retaining both the snapshot formulation and the preparation bound consists of deterministic syndrome-conditioned Pauli-$X$ recoveries of the form in Eq.~\eqref{eq:recovery}. The correction map, update schedule, and any tie-breaking rules are fixed before comparing states. Every syndrome is retained without postselection, and the same correction determines both decoded loop values. These conditions preserve the multiplicativity in Eq.~\eqref{eq:decoded-product}.

The locality requirement concerns the information used to determine each $w_i(z)$. For an alternative decoder, let $r$ bound this dependence radius, including syndrome extraction and classical processing; it need not equal the number of update rounds. Each $w_i(z)$ must depend only on input spins within distance $r$ of $C_i$, with $r=o(L)$. Merely applying on-site Pauli corrections after a global computation does not establish this bound. Other bounded-range update rules or predetermined directional schedules are admissible when their total dependence radius satisfies it. The proof in Sec.~\ref{sec:preparation} then applies unchanged whenever $2(R+r)<s$, where $R$ is the preparation circuit's causal radius. Neither successful removal of all defects nor a growing $r$ is required for this obstruction. Leaving defect-free configurations unchanged, as Sweep does, additionally preserves the reference examples above.

Useful decoding performance is a separate requirement. To obtain $f_W\to1$ for a chosen Gibbs family, an alternative decoder must yield $P_{\rm fail}\to0$ and $M_i\to0$, as follows from Eqs.~\eqref{eq:residual} and~\eqref{eq:covariance-bound}. Locality alone guarantees neither these limits nor a decoder-independent $1/0$ classification. The values and crossover regions reported below refer specifically to the chosen Sweep rule and depth; a different decoder requires its own performance analysis. Section~\ref{sec:phase-invariant} optimizes over a common class of tube observables and physical recoveries, and uses an asymptotic radius prescription to construct a phase invariant. The decoded observables of the present deterministic class enter that optimization with identity recovery.

\section{Obstruction to local preparation}\label{sec:preparation}

We next prove that disjoint preparation and decoding neighborhoods prevent correlations from product inputs, then extend this obstruction to approximate and quasi-local preparation.

\subsection{Exact preparation by local circuits}
Consider a product input $\pi\coloneqq\bigotimes_\ell\pi_\ell$ and a preparation channel $\Phi$ implemented by geometrically local gates with initially product ancillas. If the gate diameter is at most $a$ and the circuit depth is $d$, its causal radius obeys $R\leq ad$. We allow $R=o(L)$, including constant-depth and polylogarithmic-depth circuits with bounded-range gates. Our basic claim is that the prepared state $\sigma\coloneqq\Phi(\pi)$ satisfies
\begin{equation}
2(R+r)<s\quad\Longrightarrow\quad f_W(\sigma)=0.
\label{eq:causal}
\end{equation}

The input may be pure or mixed, as in the strong-field and infinite-temperature limits. At fixed finite size and temperature, taking $h\coloneqq\sqrt{h_x^2+h_z^2}\to\infty$ at fixed field direction gives a pure product state with every spin polarized along $(h_x,0,h_z)/h$. At fixed Hamiltonian, \mbox{$T\to\infty$} gives the mixed product state
\begin{equation}
\rho_\infty\coloneqq2^{-3L^3}\id=\bigotimes_\ell\frac{\id_\ell}{2}.
\label{eq:infinite-temperature}
\end{equation}
Thus Eq.~\eqref{eq:causal} applies to preparations from either limit.

To prove Eq.~\eqref{eq:causal}, we realize the channel as a local unitary circuit $U$ on the system and ancillas, with full product input
\begin{equation}
\Pi\coloneqq\pi\otimes\pi_{\rm anc},\qquad
\sigma=\Tr_{\rm anc}(U\Pi U^\dagger).
\label{eq:unitary-dilation}
\end{equation}
The input observables
\begin{equation}
O_i\coloneqq U^\dagger(\widetilde W_i\otimes\id)U
\label{eq:input-observables}
\end{equation}
occupy radius-$(R+r)$ tubes around $C_i$, as illustrated in Fig.~\ref{fig:loop-geometry}(b). Under the condition in Eq.~\eqref{eq:causal}, their supports are disjoint. Using $\widetilde W_{12}=\widetilde W_1\widetilde W_2$ and product-input factorization gives
\begin{equation}
\begin{aligned}
D_{12}(\sigma)&=\Tr(\Pi O_1O_2)=\Tr(\Pi O_1)\Tr(\Pi O_2)\\
&=M_1(\sigma)M_2(\sigma).
\end{aligned}
\label{eq:factorization}
\end{equation}
This proves Eq.~\eqref{eq:causal}, including after output ancillas are discarded. The product is preserved by unitary conjugation in the enlarged system; no multiplicativity of $\Phi^\dagger$ is assumed. A globally shared random bit is excluded by the product-ancilla hypothesis. Since $s\sim L/2$ and $r,R=o(L)$, the separation condition holds for all sufficiently large $L$.

\subsection{Approximate preparation and quasi-local evolution}

The bound also constrains approximate preparation. For a target state $\rho$ on the size-$L$ lattice, define the preparation error
\begin{equation}
\delta\coloneqq\|\rho-\sigma\|_1,
\label{eq:preparation-distance}
\end{equation}
where ${\|A\|_1\coloneqq\Tr\sqrt{A^\dagger A}}$ is the trace norm. We write $\|A\|$ for the operator norm, the largest singular value of $A$. Each decoded observable has operator norm at most one, so every moment changes by at most $\delta$. Adding and subtracting $M_1(\sigma)M_2(\rho)$ in the difference of the one-point products, and using $|M_i|\leq1$, gives
\begin{equation}
\begin{aligned}
|f_W(\rho)-f_W(\sigma)|
&\leq |D_{12}(\rho)-D_{12}(\sigma)|\\
&\quad+\sum_{i=1}^2|M_i(\rho)-M_i(\sigma)|\\
&\leq3\delta.
\end{aligned}
\label{eq:trace-stability}
\end{equation}
For a state prepared under the conditions of Eq.~\eqref{eq:causal}, this implies $|f_W(\rho)|\leq3\delta$. Thus approximate preparation cannot evade the exact obstruction by making a vanishing trace-norm error: a nonzero asymptotic value imposes a nonvanishing error.

We now extend the circuit result, Eq.~\eqref{eq:causal}, to continuous-time Hamiltonian or Lindblad evolution $\sigma=\Phi(\pi)$, whose propagation tails require error control. Enlarge each decoded tube by a buffer $b$, keeping $2(r+b)<s$. Construct the cutoff generator by retaining only Hamiltonian terms and Lindblad jump operators supported wholly within one individual tube, omitting every term that crosses a tube boundary or couples the tubes. The resulting physical channel $\Phi^{\rm cut}$ has a product-input output that factorizes between the tubes, so its decoded covariance is exactly zero. For $j=1,2,12$, define the truncation errors
\begin{equation}
\varepsilon_j\coloneqq\bigl\|[\Phi^\dagger-(\Phi^{\rm cut})^\dagger](\widetilde W_j)\bigr\|.
\label{eq:quasilocal-errors}
\end{equation}
Applying the same factorization and telescoping to these three moments yields
\begin{equation}
|f_W(\rho)|\leq\varepsilon_{12}+\varepsilon_1
+\varepsilon_2+3\delta.
\label{eq:preparation-error}
\end{equation}
The same cutoff channel controls all three moments. For Hamiltonian or local Lindblad evolution with uniformly bounded interaction range and strength per site, Lieb--Robinson truncation bounds give~\cite{liebRobinson1972,bravyiHastingsVerstraete2006,barthel2012quasilocality}
\begin{equation}
\sum_{j=1,2,12}\varepsilon_j
\leq C\,\operatorname{poly}(L,r,b)e^{-\mu(b-v\tau)}.
\label{eq:preparation-tail}
\end{equation}
Here $\tau$ is the physical evolution time; the constants $C,\mu>0$ and $v\geq0$, as well as the polynomial's degree and coefficients, are size independent. The polynomial includes tube-volume factors scaling as $L(1+r)^2$. Choosing $b$ a sufficiently small fixed fraction of $L$ makes this bound vanish for $r,\tau=o(L)$. Together with $\delta\to0$, Eq.~\eqref{eq:preparation-error} proves $f_W\to0$. Under these locality conditions, evolution for a time sublinear in $L$ therefore cannot prepare a state with nonzero asymptotic $f_W$ from a product input with vanishing trace-norm error. For this continuous-time extension, the correlation vanishes in the thermodynamic limit; small finite-size correlations from the propagation tails remain allowed.

\begin{figure*}[t]
\centering
\includegraphics[width=\textwidth]{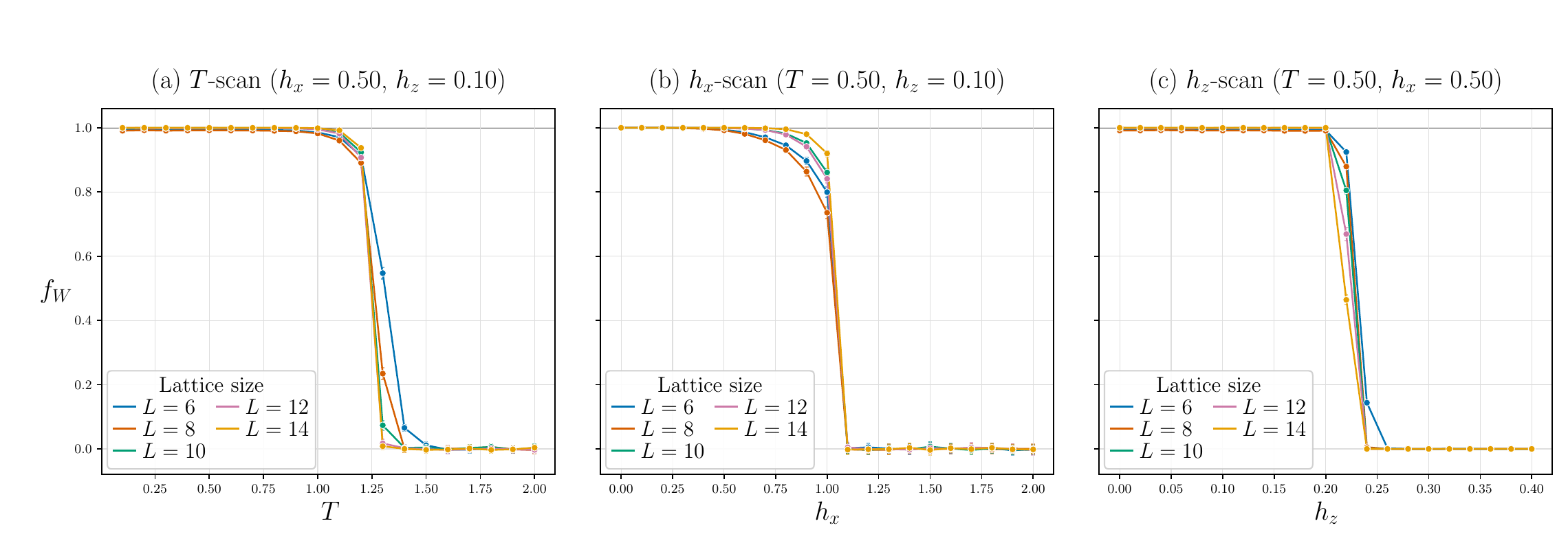}
\caption{\label{fig:scans}
Decoded Wilson-loop correlation $f_W$ for $L=6,8,10,12,14$, using the fixed decoder of Eq.~\eqref{eq:depth}. (a) Temperature scan at $(h_x,h_z)=(0.5,0.1)$. (b) Transverse-field scan at $(T,h_z)=(0.5,0.1)$. (c) Longitudinal-field scan at $(T,h_x)=(0.5,0.5)$. Error bars show nominal panelwise simultaneous $95\%$ intervals; lines guide the eye.}
\end{figure*}

\begin{figure*}[t]
\centering
\includegraphics[width=\textwidth]{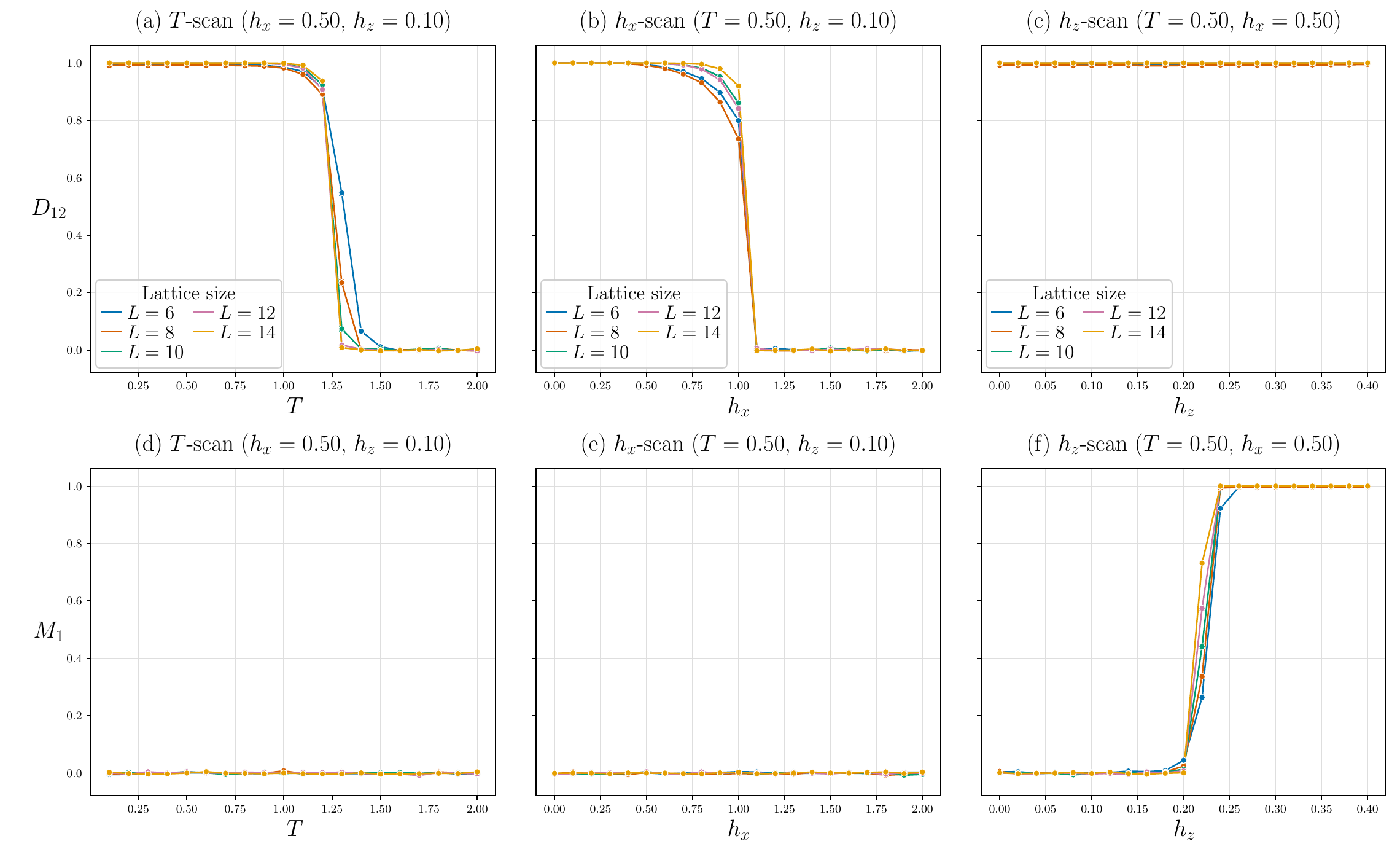}
\caption{\label{fig:scan-moments}
Sample estimates of the moments underlying $f_W$ in Fig.~\ref{fig:scans}, using the same parameter cuts, system sizes, snapshots, and decoder. (a--c) Joint moment $D_{12}$. (d--f) One-point function $M_1$. The $M_2$ estimates (not shown) nearly coincide with those of $M_1$ and are visually indistinguishable at the scale shown. The columns scan $T$, $h_x$, and $h_z$, respectively. Error bars show nominal pointwise $95\%$ percentile bootstrap intervals from whole-trajectory resampling; lines guide the eye.}
\end{figure*}

\begin{figure*}[t]
\centering
\includegraphics[width=\textwidth]{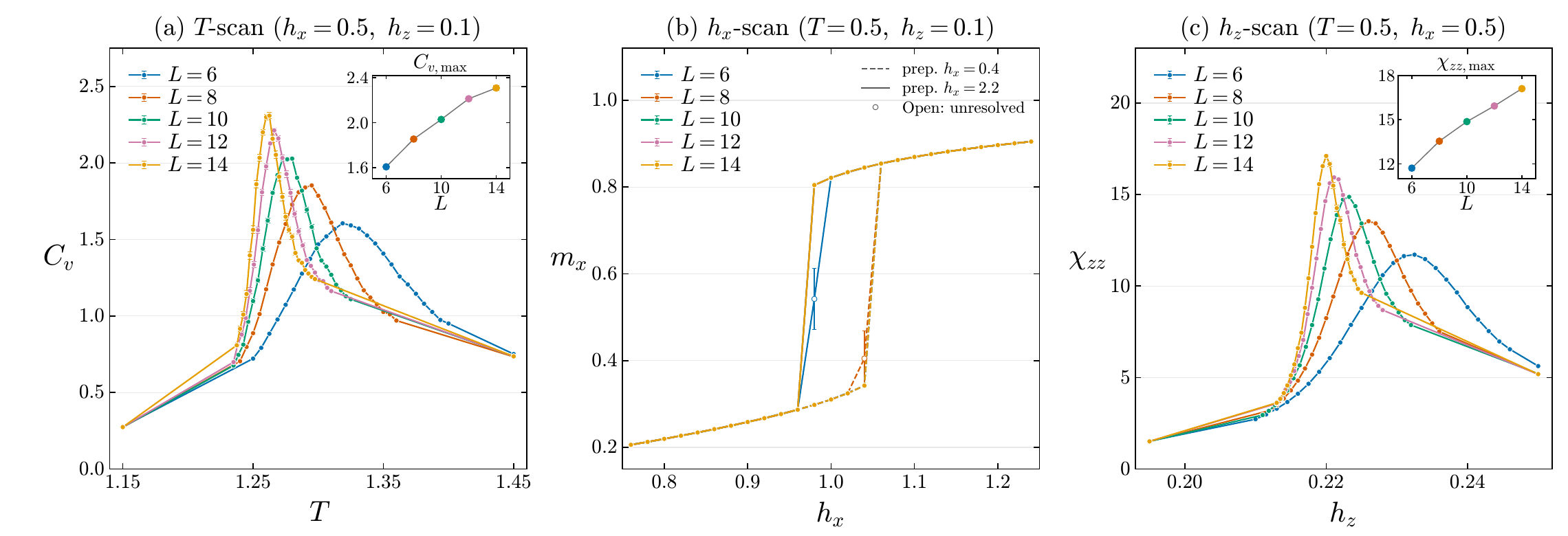}
\caption{\label{fig:thermodynamics}
Thermodynamic signatures along the cuts of Fig.~\ref{fig:scans}. (a) Specific heat per link $C_v$. (b) Transverse polarization per link $m_x$ for $L=6,8,10,12,14$. Dashed and solid curves use preparations at $h_x=0.4$ and $2.2$, respectively, with four or eight independent seeds per branch point. Error bars show the larger of the standard error across seed means and the within-seed block standard error. Open symbols mark unresolved $m_x$ chain or block diagnostics. (c) Static longitudinal susceptibility per link $\chi_{zz}$. Colors denote the same five sizes in all panels. In (a,c), error bars show fixed-block jackknife standard errors, and insets show maxima on the sampled grids with the errors of the selected points. All standard errors are provisional; lines guide the eye.}
\end{figure*}

\begin{figure*}[t]
\centering
\includegraphics[width=\textwidth]{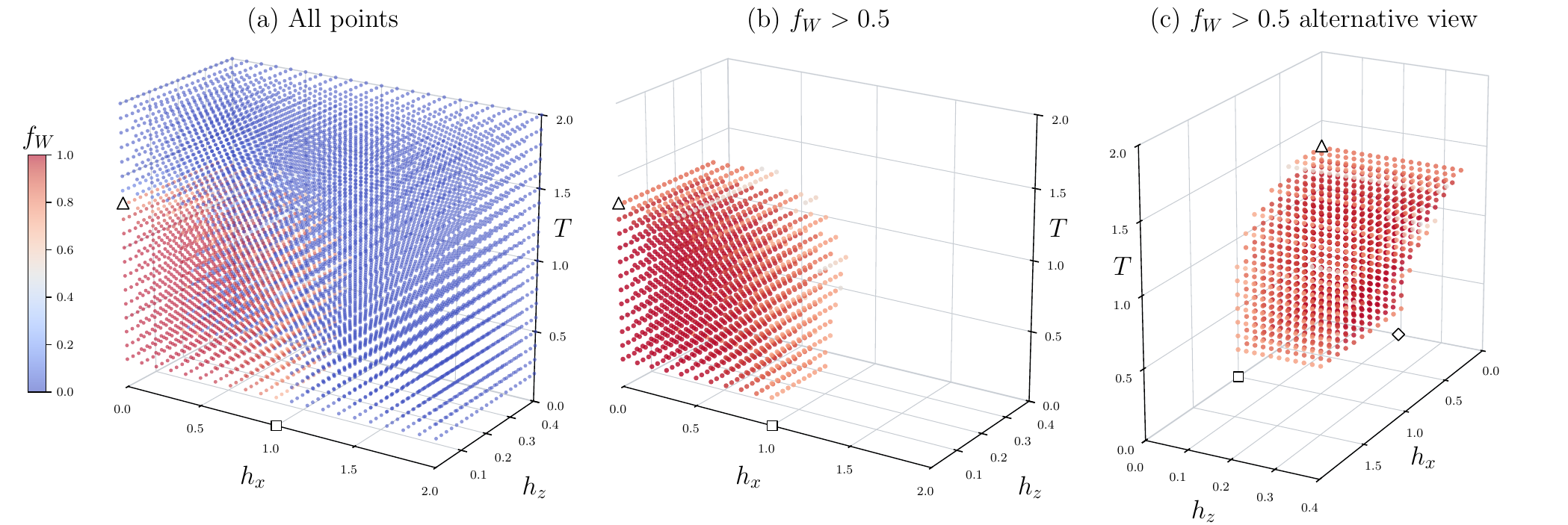}
\caption{\label{fig:cube}
Decoded Wilson-loop correlation $f_W$ at $L=8$, using the same decoder and depth prescription as Fig.~\ref{fig:scans}. (a) All $8379$ estimates. (b,c) Two views of the $1607$ points with $\widehat f_W>0.5$. Known transition points are marked by $\square$, $\diamond$, and $\triangle$ at $(h_x,h_z,T)=(1,0,0)$, $(0,0.194,0)$, and $(0,0,1.3133)$, respectively.}
\end{figure*}

\subsection{Why the product-ancilla condition matters}
\label{sec:shared-randomness}

Equations~\eqref{eq:causal} and \eqref{eq:preparation-error} require a local implementation with product ancillas. Merely requiring the channel's adjoint to preserve operator supports would be insufficient. To see the difference, denote by $\ell_z(x,y,z)$ the link from $(x,y,z)$ to $(x,y,z+1)$, with integer coordinates understood periodically, and choose the closed logical $X$ membrane
\begin{equation}
V_{\rm cl}\coloneqq\prod_{x,y=0}^{L-1}X_{\ell_z(x,y,0)}.
\label{eq:closed-membrane}
\end{equation}
These are all the $z$-directed links piercing the dual $xy$ plane at $z=1/2$, which is closed by periodicity in $x$ and $y$. Each plaquette contains zero or two of these links, so $V_{\rm cl}$ preserves the syndrome. Each loop $C_i$ contains exactly one of them, so both bare Wilson operators change sign. Since the unchanged syndrome selects the same correction, both decoded Wilson-loop operators also change sign. Now consider
\begin{equation}
\Phi_{\rm coin}(\rho)\coloneqq\tfrac12\rho+\tfrac12V_{\rm cl}\rho V_{\rm cl}^\dagger.
\label{eq:sharedcoin}
\end{equation}
Its adjoint does not enlarge any operator's support. Yet it sends the all-$Z=+1$ product state to an equal mixture of two zero-syndrome states with $(w_1,w_2)=(+,+)$ and $(-,-)$, giving $f_W=1$. The globally shared coin cannot be supplied by product ancillas in a circuit of sublinear causal radius. Each unitary branch separately preserves covariance, but their mixture does not: covariance is nonlinear in the state. Nor does the preparation theorem say that every correlated state with $f_W=0$ remains in that zero set under local operations; its hypothesis concerns preparation from a product state.

\subsection{Mixed-state phase distinction and the proposed limits}

For any product reference $\pi\coloneqq\bigotimes_\ell\pi_\ell$, with pure or mixed factors, the reset channel acts on any system operator $A$ as
\begin{equation}
\mathcal E_{\rm reset}(A)\coloneqq\Tr(A)\pi.
\label{eq:onsite-reset}
\end{equation}
It maps every density matrix to $\pi$ through a depth-one circuit of on-site reset channels: swap each system spin with an ancilla prepared in $\pi_\ell$, then discard the ancillas. All swaps act in parallel with product ancillas.

This reset construction makes all product states, including the infinite-temperature and infinite-field limits, mutually convertible and places them in the same trivial mixed-state phase. For a target $\rho$ with $\limsup_{L\to\infty}|f_W(\rho)|>0$, the preparation bounds in Eqs.~\eqref{eq:causal} and~\eqref{eq:preparation-error} exclude preparation from any product state with vanishing trace-norm error under the stated locality assumptions. Thus $\rho\to\pi$ is possible by resetting, whereas $\pi\to\rho$ is obstructed, ruling out two-way channel equivalence. This phase distinction allows arbitrary mixed-state intermediates and requires neither a Gibbs form nor quantization or invariance of the fixed $f_W$. Establishing a nonzero thermodynamic value for the Gibbs states below remains a separate numerical question; $f_W=0$ alone does not establish local preparability.

The proposed $1/0$ limits have a decoding interpretation. Equation~\eqref{eq:residual} gives
\begin{equation}
|1-f_W|\leq2P_{\rm fail}+|M_1M_2|.
\label{eq:limitbound}
\end{equation}
Agreement with vanishing bias therefore gives $f_W\to1$. For example, a uniform bound $P_{\rm fail}\leq c_0L^3e^{-\mu r}$ with size-independent constants $c_0,\mu>0$, together with $M_i\to0$, would suffice for our depth prescription. Such a bound must control correlated Gibbs fluctuations; an independent-error decoding threshold~\cite{wang2003confinement} does not establish it. The zero limit instead follows if the joint distribution of $w_1,w_2$ factorizes, as it does for the preparations in Eq.~\eqref{eq:causal}, or if both loop values become fully polarized. The numerical tests below distinguish these mechanisms while testing the conjectured limits.

\section{Numerical results}
\label{sec:fw-numerics}

We test the proposed $1/0$ limits along three parameter cuts, compare them with thermodynamic responses, and map the three-dimensional extent of the decoded correlation.

\subsection{Quantum sampling and the decoded correlation}

We use continuous-time worldline quantum Monte Carlo~\cite{wu2012phase,linsel2026paratoric}. Its equal-time $Z$-basis snapshots have the Gibbs weight $\langle z|\rho|z\rangle$ in Eq.~\eqref{eq:snapshot-moments}, so the diagonal decoded observables yield full quantum expectations even at nonzero $h_x$. All decoded-correlation calculations use periodic boundaries and the same Sweep prescription, Eq.~\eqref{eq:depth}.

Hats denote sample estimates. The same snapshots determine $\widehat D_{12}$ and $\widehat M_i$, and we form $\widehat f_W=\widehat D_{12}-\widehat M_1\widehat M_2$ after pooling the retained samples from all trajectories. Pooling preserves correlations carried by different winding signs across trajectories; it cannot correct missing equilibrium weights. We therefore compare independent runs started from both signs and test their sampling consistency. The numerical implementation, convergence checks, and uncertainty estimates are specified in Sec.~\ref{sm:numerics} of the Supplemental Material (SM).\footnote{See Supplemental Material for the quantum Monte Carlo protocols, convergence checks, uncertainty estimates, and exploratory entropy and Fredenhagen--Marcu diagnostics. It includes Refs.~\cite{sandvik2010computational,lahiri2003resampling,vehtari2021rank,kitaev2006topological,levin2006detecting,grover2011entanglement,dong2008topological,castelnovo2007entanglement,flammia2009topological,helmes2015renyi,zou2016spurious,williamson2019spurious,kim2023universal,levin2024physical,lu2020detecting,wang2025analog,hastings2010measuring,humeniuk2012quantum,melko2010finite,isakov2011topological,zhao2022measuring,zhao2022scaling,fredenhagenMarcu83,fredenhagenMarcu86,gregor2011diagnosing,alles2025FM}.}

\subsection{Size dependence along three parameter cuts}
\label{sec:fw-cuts}

Figure~\ref{fig:scans} tests the conjectured thermodynamic $1/0$ limits for $L=6,8,10,12,14$. The decoded correlation is close to one at weak fields and low temperature and is suppressed toward zero as temperature or either field increases. The connected subtraction reveals two different ways in which this suppression occurs.

Figure~\ref{fig:scan-moments} resolves these mechanisms by displaying $D_{12}$ in the upper row and $M_1$ in the lower row, using the same snapshots as Fig.~\ref{fig:scans}. The two loops are related by a translation, and both the periodic Gibbs state and the Sweep prescription respect translations. Thus their exact equilibrium expectations satisfy $M_1=M_2$, and we show only $M_1$. The finite-sample estimates need not coincide; the estimator retains the measured product $\widehat M_1\widehat M_2$.

In the $T$ and $h_x$ scans [Fig.~\ref{fig:scan-moments}(a,b,d,e)], both measured one-point functions remain small, $\widehat M_i\simeq0$, so $\widehat f_W\simeq\widehat D_{12}$. The decrease in $\widehat f_W$ reflects a loss of agreement between the decoded loop values $w_1,w_2$.

In the $h_z$ scan [Fig.~\ref{fig:scan-moments}(c,f)], $\widehat M_i$ become appreciable and approach one while $\widehat D_{12}$ remains close to one. Here $w_1,w_2$ continue to agree but become polarized: subtracting $\widehat M_1\widehat M_2$ is essential to distinguish that agreement from a shared fluctuating winding bit. The joint moment alone would miss this suppression of the connected correlation.

\subsection{Thermodynamic signatures along the same cuts}
\label{sec:thermodynamic-scans}

Figure~\ref{fig:thermodynamics} compares the loss of decoded correlations with thermodynamic observables from separate simulations along the same parameter cuts. Define the polarizations per link by $m_x\coloneqq(3L^3)^{-1}\sum_\ell\langle X_\ell\rangle_\rho$ and $m_z\coloneqq(3L^3)^{-1}\sum_\ell\langle Z_\ell\rangle_\rho$. The response functions are
\begin{equation}
\begin{aligned}
C_v&\coloneqq\frac{1}{3L^3}\left.\frac{\partial\langle H\rangle_\rho}{\partial T}\right|_{h_x,h_z},\\
\chi_{zz}&\coloneqq\left.\frac{\partial m_z}{\partial h_z}\right|_{T,h_x}.
\end{aligned}
\label{eq:thermodynamic-responses}
\end{equation}
Here $C_v$ is the specific heat per link and $\chi_{zz}$ the static longitudinal susceptibility per link. The quantum Monte Carlo estimators, sampling budgets, and uncertainties are specified in Sec.~\ref{sm:numerics} of the SM.

The temperature and longitudinal-field scans show peaks in $C_v$ and $\chi_{zz}$, respectively, in the parameter ranges where $f_W$ decreases. The sampled peak heights increase over all five sizes, as shown in the insets of Fig.~\ref{fig:thermodynamics}(a,c). This growth supports a thermodynamic-transition interpretation of the loss of decoded correlations. The growing response peaks are compatible with continuous transitions in the three-dimensional Ising universality class, although the peak data alone establish neither the transition order nor the universality class. The insets display maxima on finite sampling grids; they are not extrapolations of the peak heights or locations to the thermodynamic limit.

The transverse-field scan in Fig.~\ref{fig:thermodynamics}(b) compares preparations at $h_x=0.4$ and $2.2$, with $(T,h_z)=(0.5,0.1)$ and $L=6,8,10,12,14$. For all five sizes, the branches remain separated near the decrease of $f_W$, while curves for different sizes largely overlap away from the sharp changes. This preparation dependence is consistent with metastability on the simulated time scales. At fixed $T,h_z$, equilibrium $m_x$ is the negative $h_x$ derivative of the free energy per link, $-T\log Z/(3L^3)$; a jump would therefore produce a thermodynamic free-energy kink. The sharp changes and preparation dependence suggest a first-order transition, but unresolved relative equilibrium weights prevent determination of an equilibrium jump, coexistence field, or transition order. The peaks and branches provide complementary evidence near the changes in $f_W$; establishing the limiting singularities and their coincidence with those changes requires finite-size scaling and controlled equilibrium sampling.

\subsection{Three-dimensional extent and scope of the evidence}
\label{sec:fw-cube}

Figure~\ref{fig:cube} maps $f_W$ over $(T,h_x,h_z)$ at $L=8$. Large correlations occupy a region of weak fields and low temperature; panels (b,c) display the points with $\widehat f_W>0.5$. This map complements the size dependence in Fig.~\ref{fig:scans}, but a thresholded map at one size does not locate a thermodynamic phase boundary. The data test the proposed $1/0$ limits without establishing their exact values or critical points.

\section{Decoder dependence and regular Gibbs paths}
\label{sec:readout-transformations}

This section explains why a fixed decoded correlation need not be a phase invariant. We exhibit a change of $f_W$ along a Gibbs path generated by on-site rotations, then recover the signal by adapting the observables and decoder. The preparation obstruction in Sec.~\ref{sec:preparation} remains valid for any one admissible choice with a nonzero asymptotic value. To obtain a quantity constant throughout a channel phase, Sec.~\ref{sec:phase-invariant} will optimize over the allowed choices.

\subsection{A fixed decoder under on-site rotations}
\label{sec:gibbs-example}

A fixed decoded correlation can change even along a path of depth-one on-site unitaries. Starting from the field-free Hamiltonian $H_0\coloneqq-\sum_v A_v-\sum_p B_p$, define at fixed $\beta>0$
\begin{equation}
\begin{aligned}
U(t)&\coloneqq\prod_\ell e^{-i\pi tY_\ell/2},\\
H(t)&\coloneqq U(t)H_0U(t)^\dagger,\\
\rho(t)&\coloneqq\frac{e^{-\beta H(t)}}{Z(t)},\qquad 0\leq t\leq1,
\end{aligned}
\label{eq:isopath}
\end{equation}
where $Y_\ell$ is the Pauli $y$ operator and $Z(t)\coloneqq\Tr e^{-\beta H(t)}$. This uniform rotation preserves translations, interaction supports and norms, and the spectrum. Thus $\rho(t)=U(t)\rho(0)U(t)^\dagger$ stays in the same channel phase, with unchanged partition function and free energy.

With the original Sweep decoder and depth prescription fixed, $f_W(t)\coloneqq f_W(\rho(t))$ is continuous at finite $L$ and satisfies
\begin{equation}
\begin{gathered}
f_W(1-t)=f_W(t),\\
f_W(1/2)=0,\qquad f_W(1)=f_W(0).
\end{gathered}
\label{eq:midpoint}
\end{equation}
At fixed $L$, the canonical zero-temperature limit has $f_W(0)=1$, so sufficiently large finite $\beta$ gives a nonzero initial value that changes to zero and back.

For the reflection and endpoint identities, $Q_Z\coloneqq\prod_\ell Z_\ell$ commutes with $H_0$ and the decoded observables and sends $U(t)$ to $U(-t)$, making $f_W$ even in $t$. Also, $U(1)$ reverses both $X_\ell$ and $Z_\ell$; each star and plaquette contains an even number of links, so $\rho(t+1)=\rho(t)$.

To prove the midpoint identity, note that
\begin{equation}
H(1/2)=-\sum_v\prod_{\ell\ni v}Z_\ell
        -\sum_p\prod_{\ell\in\partial p}X_\ell.
\label{eq:uniform-midpoint-H}
\end{equation}
These terms commute, and only identity products of the plaquette-$X$ operators contribute to diagonal matrix elements. The snapshot distribution is therefore
\begin{equation}
p_{1/2}(z)\coloneqq\langle z|\rho(1/2)|z\rangle
\propto\exp\!\left(\beta\sum_v\prod_{\ell\ni v}z_\ell\right).
\label{eq:uniform-midpoint-distribution}
\end{equation}
Expand each of $\widetilde W_1$, $\widetilde W_2$, and $\widetilde W_1\widetilde W_2$ in products $Z_F\coloneqq\prod_{\ell\in F}Z_\ell$ over link sets $F$. A star $A_v$ changes neither the syndrome nor a closed Wilson loop, so these observables commute with every $A_v$ and each contributing $F$ has even degree at every vertex. Their supports lie in the band of link anchors within periodic distance $r$ of $y=0$.

Expanding the weight in Eq.~\eqref{eq:uniform-midpoint-distribution} shows that $Z_F$ has nonzero expectation only if $F$ consists of links joining different binary vertex labels $q_v\in\{0,1\}$. Even degree then requires
\begin{equation}
\sum_{w:\,w\sim v}q_w=0\pmod2,
\label{eq:uniform-layer-recurrence}
\end{equation}
where $w\sim v$ denotes the six nearest neighbors; the six copies of $q_v$ cancel modulo two. Since $2r+1<L$, some entire anchor layer lies outside the band. Its $x$- and $z$-directed links force $q_v$ to be constant on that layer, and its $+y$ links impose the same constant on the next. Equation~\eqref{eq:uniform-layer-recurrence} propagates this constant to every layer, making $F$ empty. Thus all three decoded moments equal their maximally mixed values. The two decoded operators have disjoint supports, so their covariance is zero, proving $f_W(1/2)=0$ for every finite $\beta$.

The rotated Hamiltonians generally lie outside the two-field family in Eq.~\eqref{eq:H}, but retain locality and any initial regularity of the free energy. At fixed positive temperature, the thermodynamic values at $t=0,1/2,1$ are $c,0,c$ if $f_W(0)\to c\ne0$. The finite-size argument alone establishes neither this premise nor continuity of the limiting function.

\subsection{Adapting the loop observables and recovery}
\label{sec:adapted-protocols}

The lost signal can be recovered by adapting the protocol to the state. Write $W\coloneqq(W_1,W_2)$ and $f_W(\rho;\mathcal C)$ to make the observable pair and recovery explicit. For $\rho'\coloneqq U\rho U^\dagger$, applying $U^\dagger$ before the original recovery returns the output to $\mathcal C(\rho)$. The causal radii of the inverse circuit and recovery add, preserving $o(L)$ locality when both radii are sublinear. This adapted recovery need not be syndrome-only, and the construction does not extend automatically to irreversible channels.

Equivalently, for the path in Sec.~\ref{sec:gibbs-example}, rotate both the observables and recovery:
\begin{equation}
\begin{aligned}
W_i(t)&\coloneqq U(t)W_iU(t)^\dagger,\\
\mathcal C_t(\sigma)&\coloneqq
U(t)\mathcal C\bigl(U(t)^\dagger\sigma U(t)\bigr)U(t)^\dagger.
\end{aligned}
\label{eq:rotated-candidate}
\end{equation}
Writing $W(t)\coloneqq(W_1(t),W_2(t))$, conjugation of the recovered state and observables by the same unitary gives
\begin{equation}
f_{W(t)}(\rho(t);\mathcal C_t)
=f_W(\rho(0);\mathcal C).
\label{eq:rotated-fw}
\end{equation}
At the midpoint, $W_i(1/2)=\prod_{\ell\in C_i}X_\ell$, with consistently rotated syndrome measurements and corrections. Thus the zero of the fixed $Z$-basis diagnostic does not imply loss of recoverable winding information.

This example motivates optimizing over observables and recoveries. For general local channels, the optimization must also accommodate the increase in causal radius under composition.

\section{An invariant of mixed-state phases}
\label{sec:phase-invariant}

We now optimize the decoded correlation over observables and physical recoveries, with radius bounds and limits that accommodate channel composition. We prove that the resulting quantity is constant within the specified channel phases, then relate its exact reference values to the numerical evidence.

\subsection{Admissible observables, recoveries, and the asymptotic limit}
\label{sec:optimized-diagnostic}

Fix $0<\kappa<1/2$ and a state-independent family of winding-loop pairs with separation $s\geq\kappa L$, including $C_1,C_2$ for all sufficiently large $L$. For $0<\epsilon<\kappa/2$, admit all triples $(W_1,W_2,\mathcal C)$ satisfying:
\begin{enumerate}
\item[(i)] Each $W_i$ is Hermitian, obeys $W_i^2=\id$, and is supported near its corresponding winding loop. It need not be a Pauli product or occupy the entire loop.
\item[(ii)] The recovery $\mathcal C$ is a completely positive, trace-preserving channel implemented by a geometrically local circuit with product ancillas independent of the input. Postselection and externally supplied shared random bits are excluded.
\item[(iii)] The bare observable width plus the causal enlargement in the recovery's physical implementation is at most $\epsilon L$, including all ancillary and classical processing. The two full backward cones are therefore disjoint.
\end{enumerate}
Identity recovery is allowed. The geometry and resource bounds are fixed when comparing states; only the candidate chosen within these bounds may depend on the state. Define the finite-size optimum directly by
\begin{equation}
F_{\epsilon,L}(\rho_L)
\coloneqq\sup_{(W_1,W_2,\mathcal C)}
|f_W(\rho_L;\mathcal C)|,
\label{eq:radius-optimized-fw}
\end{equation}
where the supremum runs over all triples satisfying (i)--(iii). The binary covariance bound gives $0\leq F_{\epsilon,L}\leq1$. On-site unitaries preserve this optimum because Eq.~\eqref{eq:rotated-candidate} bijectively maps the admissible candidates to themselves without enlarging their supports.

A general local channel can increase the required radius. To absorb any $o(L)$ increase, define for a state family $\boldsymbol\rho\coloneqq\{\rho_L\}_L$
\begin{equation}
\mathcal{F}(\boldsymbol\rho)
\coloneqq\lim_{\epsilon\downarrow0}\limsup_{L\to\infty}
F_{\epsilon,L}(\rho_L).
\label{eq:phase-invariant}
\end{equation}
The upper limit avoids assuming convergence in $L$; the outer limit exists because the bounded optima decrease as the radius budget shrinks. Taking $L\to\infty$ first leaves room for any sublinear enlargement, after which $\epsilon\downarrow0$ makes the allowed radius fraction vanish. The chosen geometric family remains fixed throughout.

\subsection{Local-channel monotonicity and phase invariance}
\label{sec:phase-circuit-proof}

\begin{proposition}[Channel-phase invariant]
\label{prop:phase-invariant}
For the fixed geometric family above, $\mathcal{F}$ is nonincreasing under local circuit channels with product ancillas and causal radius $R_L=o(L)$, including conversions with vanishing trace-norm error. It is therefore constant under two-way conversions of this kind.
\end{proposition}

\begin{proof}
Suppose $\Phi_L$ maps $\rho_L$ to $\sigma_L$ with error $e_L\coloneqq\|\sigma_L-\Phi_L(\rho_L)\|_1\to0$. An output candidate can be applied to the input by composing its recovery with $\Phi_L$: $f_W(\Phi_L(\rho_L);\mathcal C)=f_W(\rho_L;\mathcal C\circ\Phi_L)$. This remains a physical circuit with product ancillas and enlarges the radius by at most $R_L$.

Equation~\eqref{eq:trace-stability} and trace-norm contractivity under recovery bound the effect of the conversion error by $3e_L$, uniformly over candidates. Fix $0<\epsilon<\kappa/4$. For sufficiently large $L$, $R_L/L<\epsilon$, so
\begin{equation}
\begin{aligned}
F_{\epsilon,L}(\sigma_L)
&\leq F_{\epsilon+R_L/L,L}(\rho_L)+3e_L\\
&\leq F_{2\epsilon,L}(\rho_L)+3e_L.
\end{aligned}
\label{eq:radius-monotonicity}
\end{equation}
Taking the upper limit in $L$ and then $\epsilon\downarrow0$ gives
\begin{equation}
\mathcal{F}(\boldsymbol\sigma)
\leq \mathcal{F}(\boldsymbol\rho).
\label{eq:phase-monotonicity}
\end{equation}
An allowed reverse conversion gives the opposite inequality. Thus, writing $\sim$ for two-way channel equivalence,
\begin{equation}
\boldsymbol\rho\sim\boldsymbol\sigma
\quad\Longrightarrow\quad
\mathcal{F}(\boldsymbol\rho)=\mathcal{F}(\boldsymbol\sigma).
\label{eq:phase-equality}
\end{equation}
\end{proof}

For a product input, disjoint backward cones make every admissible covariance vanish. Hence $\mathcal{F}=0$ for all product-state families and, by monotonicity, all families locally preparable from them. A positive value excludes the trivial phase; the converse need not hold. Equal values need not identify the same phase, so the invariant is not a complete classification.

\subsection{Controlled quasi-local evolution}\label{sec:phase-quasilocal}

To extend the proof to Hamiltonian and Lindblad evolution, first eliminate recovery from the optimization. For any admissible candidate, $A_i\coloneqq\mathcal C^\dagger(W_i)$ are Hermitian contractions, $-\id\leq A_i\leq\id$, in the radius-$\epsilon L$ tubes. Disjoint backward cones in a unitary dilation with product ancillas give $\mathcal C^\dagger(W_1W_2)=A_1A_2$; disjoint operator supports alone would not suffice. Each contraction is a convex combination of Hermitian involutions on the same tube, obtained by expressing its eigenvalues as averages of signs $\pm1$. Bilinearity of covariance then bounds its absolute value by the largest absolute covariance of these binary pairs. Conversely, all such pairs are admissible with identity recovery. Thus $F_{\epsilon,L}$ equals the supremum of absolute input-state covariances over binary observables in those tubes.

Let $\Phi_L$ now be evolution for time $\tau_L=o(L)$ under the uniformly bounded finite-range Hamiltonian or local Lindblad generators of Eq.~\eqref{eq:preparation-tail}. For $0<\epsilon<\kappa/4$, cut the generator to two disjoint tubes of radius $2\epsilon L$. The cutoff evolution pulls binary observables in the inner tubes back to contractions in the outer tubes and preserves their joint product. The argument above bounds their absolute covariance by $F_{2\epsilon,L}(\rho_L)$. The buffer of width $\epsilon L$ gives uniform Lieb--Robinson errors for all norm-one observables and tube locations, with support-volume factors at most polynomial in $L$. Including a vanishing conversion error $e_L$ as above gives
\begin{equation}
\begin{aligned}
F_{\epsilon,L}(\sigma_L)
&\leq F_{2\epsilon,L}(\rho_L)+3e_L\\
&\quad+c_\epsilon\operatorname{poly}(L)
e^{-\mu(\epsilon L-v\tau_L)}.
\end{aligned}
\label{eq:phase-quasilocal}
\end{equation}
Here $c_\epsilon>0$ may depend on the fixed buffer fraction, while $\mu>0$, $v\geq0$, and the polynomial bounds are uniform in $L$ and the optimized observables. The last term vanishes at fixed $\epsilon$, so the same limits prove Eqs.~\eqref{eq:phase-monotonicity} and~\eqref{eq:phase-equality} for these quasi-local conversions. Invariance along a thermodynamically regular Gibbs path additionally requires such two-way channels; free-energy analyticity alone is not used in the proof.

\subsection{Exact reference values and scope of the invariant}
\label{sec:invariant-interpretation}

The invariant has two exact reference values at zero field. Let $P_{0,L}$ project onto the eight-dimensional ground space of $H_0$ on the three-torus. The canonical zero-temperature limit at each finite size is
\begin{equation}
\rho_{0,L}\coloneqq\lim_{\beta\to\infty}\frac{e^{-\beta H_0}}{Z}
=\frac{P_{0,L}}{8}.
\label{eq:ground-gibbs-mixture}
\end{equation}
Choose the identity recovery. Equation~\eqref{eq:bare-pair} gives $W_1W_2P_{0,L}=P_{0,L}$ because every plaquette has eigenvalue $+1$ in the ground space. The logical membrane $V_{\mathrm{cl}}$ in Eq.~\eqref{eq:closed-membrane} leaves $\rho_{0,L}$ invariant and reverses both $W_i$, so their individual expectations vanish. Consequently,
\begin{equation}
\langle W_1W_2\rangle_{\rho_{0,L}}=1,\qquad
\langle W_i\rangle_{\rho_{0,L}}=0,
\label{eq:ground-wilson-moments}
\end{equation}
and $f_W(\rho_{0,L};\mathrm{id})=1$ exactly. The Sweep recovery gives the same value because it makes no correction to a zero syndrome. This argument uses the full canonical ground-space mixture; a pure state with a fixed logical $Z$ parity need not have the same connected correlation.

At $T=\infty$, the state $\rho_{\infty,L}=2^{-3L^3}\id$ is the product state in Eq.~\eqref{eq:infinite-temperature}. Combining these facts with $0\leq \mathcal{F}\leq1$ yields, at $h_x=h_z=0$,
\begin{equation}
\mathcal{F}(\boldsymbol\rho_0)=1,\qquad
\mathcal{F}(\boldsymbol\rho_\infty)=0.
\label{eq:phase-endpoints}
\end{equation}
Here $\boldsymbol\rho_0\coloneqq\{\rho_{0,L}\}_L$ and $\boldsymbol\rho_\infty\coloneqq\{\rho_{\infty,L}\}_L$. These limiting families therefore belong to distinct channel phases. The endpoint comparison alone does not establish open phases at fixed positive temperature or locate a thermal transition.

\subsection{Ferromagnetic mixtures and Ising coexistence}
\label{sec:ferromagnetic-mixtures}

To illustrate the sensitivity to conventional long-range order, consider a mixture of the two fully polarized ferromagnetic configurations with a size-independent weight $0\leq p\leq1$,
\begin{equation}
\begin{aligned}
\rho_{p,L}&\coloneqq p\pi_{0,L}+(1-p)\pi_{1,L},\\
\pi_{a,L}&\coloneqq\bigl(|a\rangle\langle a|\bigr)^{\otimes 3L^3},
\end{aligned}
\label{eq:ferromagnetic-mixture}
\end{equation}
where $a=0,1$ and $Z|a\rangle=(-1)^a|a\rangle$. Choose one link $\ell_i\in C_i$ in each of the two separated tubes and take $W_i\coloneqq Z_{\ell_i}$ with $\mathcal C=\mathrm{id}$. These single-spin observables are admissible because their supports lie within the permitted tubes. They give $\langle W_i\rangle_{\rho_{p,L}}=2p-1$ and $\langle W_1W_2\rangle_{\rho_{p,L}}=1$, hence $f_W(\rho_{p,L};\mathrm{id})=4p(1-p)$.

No admissible recovery can yield a larger absolute covariance. To see this, for an arbitrary candidate define the conditional means $m_i^{(a)}\coloneqq\langle W_i\rangle_{\mathcal C(\pi_{a,L})}\in[-1,1]$. Within either polarized component, the input spins and ancillas are in a product state. The disjoint backward cones therefore give $\langle W_1W_2\rangle_{\mathcal C(\pi_{a,L})}=m_1^{(a)}m_2^{(a)}$. Linearity of the channel then yields
\begin{align}
f_W(\rho_{p,L};\mathcal C)
&=p(1-p)\bigl(m_1^{(0)}-m_1^{(1)}\bigr)\bigl(m_2^{(0)}-m_2^{(1)}\bigr),\notag\\
|f_W(\rho_{p,L};\mathcal C)|&\leq4p(1-p).
\label{eq:ferromagnetic-bound}
\end{align}
This bound holds uniformly over all candidates and is attained by the single-spin choice above. Thus $F_{\epsilon,L}(\rho_{p,L})=4p(1-p)$, and for $\boldsymbol\rho_p\coloneqq\{\rho_{p,L}\}_L$,
\begin{equation}
\mathcal{F}(\boldsymbol\rho_p)=4p(1-p).
\label{eq:ferromagnetic-invariant}
\end{equation}
As $p$ varies, $\mathcal{F}=4p(1-p)$ spans $[0,1]$; the equal mixture has $\mathcal{F}=1$. The invariant is therefore not generally quantized. This example does not establish intermediate values for the canonical toric-code Gibbs family.

The equal mixture should be distinguished from a Gibbs state selected by a nonzero magnetic field. As a separate illustration, consider the nearest-neighbor Ising model $H_{\mathrm{Is}}\coloneqq-\sum_{\langle i,j\rangle}Z_iZ_j-h_z\sum_iZ_i$ on a periodic $L\times L$ square lattice, with the ferromagnetic coupling set to one. Here $Z_i$ is the Pauli operator at site $i$, and $\langle i,j\rangle$ denotes a nearest-neighbor pair. Its canonical $T\to0$ limit at each finite size is the equal mixture of the all-up and all-down configurations when $h_z=0$, but is the all-up product state for every fixed $h_z>0$. The same invariant construction, applied in two dimensions with separated strips around winding loops, therefore assigns $\mathcal{F}=1$ and $0$ to these respective zero-temperature state families. A temperature path at fixed positive field does not start from the equal mixture.

This distinction is consistent with the absence of a thermodynamic transition at $T>0,h_z>0$~\cite{leeYang1952ising}. Below the zero-field Ising critical temperature $T_c^{\mathrm{Is}}$, the line $h_z=0$ is a coexistence line. For $0<T<T_c^{\mathrm{Is}}$, the thermodynamic free energy per spin $f_{\mathrm{Is}}$ has the small-field behavior
\begin{equation}
f_{\mathrm{Is}}(T,h_z)=f_{\mathrm{Is}}(T,0)-m_*(T)|h_z|+o(|h_z|),
\label{eq:ising-coexistence-cusp}
\end{equation}
where $m_*(T)>0$ is the spontaneous magnetization per spin. Thus the symmetric mixture lies on a thermodynamically singular boundary of the regular positive-field region. Taking $h_z\downarrow0$ after the thermodynamic limit selects the positively magnetized phase, whereas setting $h_z=0$ before that limit gives the symmetric mixture of the two phases. The two orders of limits describe different states.

At fixed $T>0,h_z>0$, the Gibbs state is generally not a product state: its magnetization is nonzero, while its connected spin correlations decay exponentially~\cite{frohlichRodriguez2017cluster}. Establishing $\mathcal{F}=0$ for such finite-temperature families requires bounds uniform over the growing tube observables in the optimization; decay for fixed single-spin observables alone is not that bound. The ferromagnetic mixture example therefore does not imply that a unit-value region persists at nonzero field.

\subsection{Implications for the numerical results}
\label{sec:invariant-numerics}

To connect the invariant to the numerical results, recall that the Sweep-decoded operators are Hermitian involutions in radius-$r$ tubes, with $r=o(L)$. Taking those operators as the admissible tube observables and using identity recovery reproduces their covariance. Hence
\begin{equation}
\mathcal{F}(\boldsymbol\rho)
\geq\limsup_{L\to\infty}|f_W(\rho_L)|,
\label{eq:phase-sweep-bound}
\end{equation}
where the right-hand side uses the fixed Sweep prescription of the numerical calculations. In particular, $f_W\to1$ implies $\mathcal{F}=1$.

Figures~\ref{fig:scans} and~\ref{fig:cube} suggest $f_W\to1$ at weak fields and low temperature, and hence $\mathcal{F}=1$ through Eq.~\eqref{eq:phase-sweep-bound}. Suppressed fixed-decoder correlations elsewhere are consistent with $\mathcal{F}=0$, but provide no optimized upper bound: another admissible observable or recovery could retain nonzero covariance. The plots therefore do not establish a phase diagram of $\mathcal{F}$.

\section{Discussion}\label{sec:discussion}
The decoded correlation $f_W$ is accessible from quantum Monte Carlo snapshots and obstructs preparation from product inputs when its asymptotic value is nonzero. Its optimized counterpart $\mathcal{F}$ is invariant under the specified two-way local channels and distinguishes the canonical field-free ground-state mixture from product states. It also detects conventional long-range correlations and is not generally quantized.

The numerical results suggest an extended region with $\mathcal{F}=1$, while suppressed fixed-decoder correlations elsewhere are only consistent with $\mathcal{F}=0$. Growing response peaks and preparation-dependent polarization branches provide complementary thermodynamic evidence without determining the limiting singularities. Relating channel phases to thermodynamic phases additionally requires control of channels along regular Gibbs paths.

Entropy-based and Fredenhagen--Marcu diagnostics offer complementary tests; our exploratory calculations and their limitations are described in Secs.~\ref{sm:tee} and~\ref{sm:fm} of the SM.

The reference states allowed in a preparation problem also matter for its quantum interpretation. At $T>0$, the field-free three-dimensional toric-code state is circuit-trivial when correlated classical Gibbs states are allowed as inputs~\cite{hastings2011topological}. This broader input class can already carry classical correlations, so that result is compatible with the product-input obstruction proved here. Establishing long-range quantum entanglement requires additional information. Anomalous one-form symmetries can, for example, enforce such entanglement in mixed states~\cite{li2025entanglement}; the winding correlation studied here alone does not establish it.

The field-free winding variables admit a classical-memory interpretation. On the three-torus, the three independent logical $Z$ parities label $2^3=8$ configurations of a three-bit classical register. Below the thermal transition, classical sector information can remain stable while intersector coherence need not be preserved~\cite{castelnovo2008topological}; this does not imply a protected quantum memory. Our loop pair probes one of these three symmetry-related variables. At nonzero fields, establishing a three-bit memory requires reliable recovery of all eight messages and a separate analysis of their storage lifetime.

Such a lifetime is a dynamical property. In weak-coupling Markovian analyses, relaxation times of the two-dimensional toric code are bounded independently of system size~\cite{alicki2009thermalization}, whereas the four-dimensional code supports exponentially long-lived logical observables at sufficiently low temperature~\cite{alicki2010thermal}. In the classical three-dimensional $\mathbb Z_2$ gauge theory, robust finite-temperature classical memory persists under explicit one-form symmetry breaking~\cite{stahl2026slow}. This provides a useful comparison, but does not establish memory dynamics in the present noncommuting quantum model.

By connecting a rigorous invariant of phases under two-way local channels to a diagnostic accessible to quantum Monte Carlo, this work provides a concrete route to investigating Gibbs-state phases beyond microscopic symmetry breaking. The three-dimensional toric code illustrates how this approach can be applied to interacting thermal states even when both microscopic higher-form symmetries are explicitly broken.

\begin{acknowledgments}
The author thanks Claudio Castelnovo, Ryohei Kobayashi, Jong Yeon Lee, Zi Yang Meng, Yukitoshi Motome, Seishiro Ono, Tibor Rakovszky, Cenke Xu, and Liujun Zou for helpful discussions. Part of this work advanced through discussions with participants of the workshops ``Recent advances in studies of strongly correlated electrons'' (ICISE, Quy Nhon, Vietnam, June 1--5, 2026) and ``Quantum Matter Frontiers: Entanglement, Symmetry, and Emergence'' (KIAS, Seoul, June 15--19, 2026). A Hong Kong University of Science and Technology (HKUST) startup grant supported this work.
\end{acknowledgments}

\bibliography{references,references_sm}

@article{landau1937theory,
  author       = {Landau, L. D.},
  title        = {On the Theory of Phase Transitions},
  journal      = {Zh. Eksp. Teor. Fiz.},
  volume       = {7},
  pages        = {19},
  year         = {1937},
  note         = {English translation: Collected Papers of L.~D.~Landau, ed. D.~ter~Haar (Pergamon, 1965), p.~193}
}

@book{anderson1984basic,
  author       = {Anderson, P. W.},
  title        = {Basic Notions of Condensed Matter Physics},
  publisher    = {Benjamin/Cummings},
  address      = {Menlo Park, CA},
  year         = {1984}
}

@book{stanley1971introduction,
  author       = {Stanley, H. E.},
  title        = {Introduction to Phase Transitions and Critical Phenomena},
  publisher    = {Oxford University Press},
  address      = {Oxford},
  year         = {1971}
}

@article{gaiotto2015generalized,
  author       = {Gaiotto, Davide and Kapustin, Anton and Seiberg, Nathan and Willett, Brian},
  title        = {Generalized global symmetries},
  journal      = {J. High Energy Phys.},
  volume       = {2015},
  number       = {2},
  pages        = {172},
  year         = {2015},
  doi          = {10.1007/JHEP02(2015)172},
}

@article{lake2018higher,
  author       = {Lake, Ethan},
  title        = {Higher-form symmetries and spontaneous symmetry breaking},
  journal      = {\href{https://arxiv.org/abs/1802.07747}{arXiv:1802.07747}},
  year         = {2018},
}

@article{yangShiLee2025topological,
  author       = {Yang, Tai-Hsuan and Shi, Bowen and Lee, Jong Yeon},
  title        = {Topological Mixed States: Phases of Matter from Axiomatic Approaches},
  journal      = {\href{https://arxiv.org/abs/2506.04221}{arXiv:2506.04221}},
  year         = {2025},
}

@article{mcgreevy2022generalized,
  author       = {McGreevy, John},
  title        = {Generalized Symmetries in Condensed Matter},
  journal      = {Annu. Rev. Condens. Matter Phys.},
  volume       = {14},
  pages        = {57--82},
  year         = {2023},
  doi          = {10.1146/annurev-conmatphys-040721-021029},
}

@article{wegner1971duality,
  author       = {Wegner, Franz J.},
  title        = {Duality in generalized {I}sing models and phase transitions without local order parameters},
  journal      = {J. Math. Phys.},
  volume       = {12},
  number       = {10},
  pages        = {2259--2272},
  year         = {1971},
  doi          = {10.1063/1.1665530}
}

@article{fradkin1979phase,
  author       = {Fradkin, Eduardo and Shenker, Stephen H.},
  title        = {Phase diagrams of lattice gauge theories with {H}iggs fields},
  journal      = {Phys. Rev. D},
  volume       = {19},
  pages        = {3682--3697},
  year         = {1979},
  doi          = {10.1103/PhysRevD.19.3682}
}

@article{elitzur1975impossibility,
  author       = {Elitzur, S.},
  title        = {Impossibility of spontaneously breaking local symmetries},
  journal      = {Phys. Rev. D},
  volume       = {12},
  pages        = {3978--3982},
  year         = {1975},
  doi          = {10.1103/PhysRevD.12.3978}
}

@article{osterwalder1978gauge,
  author       = {Osterwalder, Konrad and Seiler, Erhard},
  title        = {Gauge field theories on a lattice},
  journal      = {Ann. Phys. (N.Y.)},
  volume       = {110},
  number       = {2},
  pages        = {440--471},
  year         = {1978},
  doi          = {10.1016/0003-4916(78)90039-8}
}

@article{berezinskii1971destruction,
  author       = {Berezinskii, V. L.},
  title        = {Destruction of long-range order in one-dimensional and two-dimensional systems possessing a continuous symmetry group. {II}. {Q}uantum systems},
  journal      = {Sov. Phys. JETP},
  volume       = {34},
  pages        = {610},
  year         = {1972},
  note         = {English translation of Zh.~Eksp.~Teor.~Fiz.~\textbf{61}, 1144 (1971)}
}

@article{kosterlitz1973ordering,
  author       = {Kosterlitz, J. M. and Thouless, D. J.},
  title        = {Ordering, metastability and phase transitions in two-dimensional systems},
  journal      = {J. Phys. C: Solid State Phys.},
  volume       = {6},
  number       = {7},
  pages        = {1181--1203},
  year         = {1973},
  doi          = {10.1088/0022-3719/6/7/010}
}

@article{mermin1966absence,
  author       = {Mermin, N. D. and Wagner, H.},
  title        = {Absence of ferromagnetism or antiferromagnetism in one- or two-dimensional isotropic {H}eisenberg models},
  journal      = {Phys. Rev. Lett.},
  volume       = {17},
  pages        = {1133--1136},
  year         = {1966},
  doi          = {10.1103/PhysRevLett.17.1133}
}

@article{kosterlitz1974critical,
  author       = {Kosterlitz, J. M.},
  title        = {The critical properties of the two-dimensional {XY} model},
  journal      = {J. Phys. C: Solid State Phys.},
  volume       = {7},
  number       = {6},
  pages        = {1046--1060},
  year         = {1974},
  doi          = {10.1088/0022-3719/7/6/005}
}

@article{hasenbusch2005,
  author       = {Hasenbusch, Martin},
  title        = {The two-dimensional {XY} model at the transition temperature: A high-precision {M}onte {C}arlo study},
  journal      = {J. Phys. A: Math. Gen.},
  volume       = {38},
  number       = {26},
  pages        = {5869--5883},
  year         = {2005},
  doi          = {10.1088/0305-4470/38/26/003}
}

@article{castelnovo2008topological,
  author       = {Castelnovo, Claudio and Chamon, Claudio},
  title        = {Topological order in a three-dimensional toric code at finite temperature},
  journal      = {Phys. Rev. B},
  volume       = {78},
  pages        = {155120},
  year         = {2008},
  doi          = {10.1103/PhysRevB.78.155120}
}

@article{castelnovo2007entanglement,
  author       = {Castelnovo, Claudio and Chamon, Claudio},
  title        = {Entanglement and topological entropy of the toric code at finite temperature},
  journal      = {Phys. Rev. B},
  volume       = {76},
  pages        = {184442},
  year         = {2007},
  doi          = {10.1103/PhysRevB.76.184442}
}

@article{kitaev2006topological,
  author       = {Kitaev, Alexei and Preskill, John},
  title        = {Topological entanglement entropy},
  journal      = {Phys. Rev. Lett.},
  volume       = {96},
  pages        = {110404},
  year         = {2006},
  doi          = {10.1103/PhysRevLett.96.110404}
}

@article{levin2006detecting,
  author       = {Levin, Michael and Wen, Xiao-Gang},
  title        = {Detecting topological order in a ground state wave function},
  journal      = {Phys. Rev. Lett.},
  volume       = {96},
  pages        = {110405},
  year         = {2006},
  doi          = {10.1103/PhysRevLett.96.110405}
}

@article{grover2011entanglement,
  author       = {Grover, Tarun and Turner, Ari M. and Vishwanath, Ashvin},
  title        = {Entanglement entropy of gapped phases and topological order in three dimensions},
  journal      = {Phys. Rev. B},
  volume       = {84},
  pages        = {195120},
  year         = {2011},
  doi          = {10.1103/PhysRevB.84.195120}
}

@article{hastings2005quasiadiabatic,
  author       = {Hastings, Matthew B. and Wen, Xiao-Gang},
  title        = {Quasiadiabatic continuation of quantum states: The stability of topological ground-state degeneracy and emergent gauge invariance},
  journal      = {Phys. Rev. B},
  volume       = {72},
  pages        = {045141},
  year         = {2005},
  doi          = {10.1103/PhysRevB.72.045141}
}

@article{bravyi2010topological,
  doi          = {10.1063/1.3490195},
  author       = {Bravyi, Sergey and Hastings, Matthew B. and Michalakis, Spyridon},
  title        = {Topological quantum order: Stability under local perturbations},
  journal      = {J. Math. Phys.},
  volume       = {51},
  pages        = {093512},
  year         = {2010}
}

@article{zou2016spurious,
  author       = {Zou, Liujun and Haah, Jeongwan},
  title        = {Spurious long-range entanglement and replica correlation length},
  journal      = {Phys. Rev. B},
  volume       = {94},
  pages        = {075151},
  year         = {2016},
  doi          = {10.1103/PhysRevB.94.075151}
}

@article{williamson2019spurious,
  author       = {Williamson, Dominic J. and Dua, Arpit and Cheng, Meng},
  title        = {Spurious topological entanglement entropy from subsystem symmetries},
  journal      = {Phys. Rev. Lett.},
  volume       = {122},
  pages        = {140506},
  year         = {2019},
  doi          = {10.1103/PhysRevLett.122.140506}
}

@article{kim2023universal,
  author       = {Kim, Isaac H. and Levin, Michael and Lin, Ting-Chun and Ranard, Daniel and Shi, Bowen},
  title        = {Universal lower bound on topological entanglement entropy},
  journal      = {Phys. Rev. Lett.},
  volume       = {131},
  pages        = {166601},
  year         = {2023},
  doi          = {10.1103/PhysRevLett.131.166601}
}

@article{levin2024physical,
  author        = {Levin, Michael},
  title         = {Physical proof of the topological entanglement entropy inequality},
  journal       = {Phys. Rev. B},
  year          = {2024},
  volume        = {110},
  pages         = {165154},
  doi           = {10.1103/PhysRevB.110.165154},
}

@article{hastings2010measuring,
  author       = {Hastings, Matthew B. and Gonz\'alez, Iv\'an and Kallin, Ann B. and Melko, Roger G.},
  title        = {Measuring {R}\'enyi entanglement entropy in quantum {M}onte {C}arlo simulations},
  journal      = {Phys. Rev. Lett.},
  volume       = {104},
  pages        = {157201},
  year         = {2010},
  doi          = {10.1103/PhysRevLett.104.157201}
}

@article{melko2010finite,
  author       = {Melko, Roger G. and Kallin, Ann B. and Hastings, Matthew B.},
  title        = {Finite-size scaling of mutual information in {M}onte {C}arlo simulations: Application to the spin-1/2 {XXZ} model},
  journal      = {Phys. Rev. B},
  volume       = {82},
  pages        = {100409(R)},
  year         = {2010},
  doi          = {10.1103/PhysRevB.82.100409}
}

@article{isakov2011topological,
  author       = {Isakov, Sergei V. and Hastings, Matthew B. and Melko, Roger G.},
  title        = {Topological entanglement entropy of a {B}ose--{H}ubbard spin liquid},
  journal      = {Nat. Phys.},
  volume       = {7},
  pages        = {772},
  year         = {2011},
  doi          = {10.1038/nphys2036}
}

@article{humeniuk2012quantum,
  author       = {Humeniuk, Stephan and Roscilde, Tommaso},
  title        = {Quantum {M}onte {C}arlo calculation of entanglement {R}\'enyi entropies for generic quantum systems},
  journal      = {Phys. Rev. B},
  volume       = {86},
  pages        = {235116},
  year         = {2012},
  doi          = {10.1103/PhysRevB.86.235116}
}

@article{pelissetto2002critical,
  author       = {Pelissetto, Andrea and Vicari, Ettore},
  title        = {Critical phenomena and renormalization-group theory},
  journal      = {Phys. Rep.},
  volume       = {368},
  number       = {6},
  pages        = {549--727},
  year         = {2002},
  doi          = {10.1016/S0370-1573(02)00219-3}
}

@article{dennis2002topological,
  author       = {Dennis, Eric and Kitaev, Alexei and Landahl, Andrew and Preskill, John},
  title        = {Topological quantum memory},
  journal      = {J. Math. Phys.},
  volume       = {43},
  number       = {9},
  pages        = {4452--4505},
  year         = {2002},
  doi          = {10.1063/1.1499754}
}

@article{wang2003confinement,
  author       = {Wang, Chenyang and Harrington, Jim and Preskill, John},
  title        = {Confinement--{Higgs} transition in a disordered gauge theory and the accuracy threshold for quantum memory},
  journal      = {Ann. Phys.},
  volume       = {303},
  pages        = {31},
  year         = {2003},
  doi          = {10.1016/S0003-4916(02)00019-2}
}

@article{kubica2019cellular,
  author       = {Kubica, Aleksander and Preskill, John},
  title        = {Cellular-automaton decoders with provable thresholds for topological codes},
  journal      = {Phys. Rev. Lett.},
  volume       = {123},
  pages        = {020501},
  year         = {2019},
  doi          = {10.1103/PhysRevLett.123.020501}
}

@article{liebRobinson1972,
  author       = {Lieb, Elliott H. and Robinson, Derek W.},
  title        = {The finite group velocity of quantum spin systems},
  journal      = {Commun. Math. Phys.},
  volume       = {28},
  pages        = {251},
  year         = {1972},
  doi          = {10.1007/BF01645779}
}

@article{bravyiHastingsVerstraete2006,
  author       = {Bravyi, S. and Hastings, M. B. and Verstraete, F.},
  title        = {{Lieb-Robinson} bounds and the generation of correlations and topological quantum order},
  journal      = {Phys. Rev. Lett.},
  volume       = {97},
  pages        = {050401},
  year         = {2006},
  doi          = {10.1103/PhysRevLett.97.050401}
}

@article{kitaev2003fault,
  author       = {Kitaev, A. Yu.},
  title        = {Fault-tolerant quantum computation by anyons},
  journal      = {Ann. Phys. (N.Y.)},
  volume       = {303},
  number       = {1},
  pages        = {2--30},
  year         = {2003},
  doi          = {10.1016/S0003-4916(02)00018-0}
}

@article{tupitsyn2010topological,
  author       = {Tupitsyn, Igor S. and Kitaev, Alexei and Prokof'ev, Nikolai V. and Stamp, Philip C. E.},
  title        = {Topological multicritical point in the phase diagram of the toric code model and three-dimensional lattice gauge {H}iggs model},
  journal      = {Phys. Rev. B},
  volume       = {82},
  pages        = {085114},
  year         = {2010},
  doi          = {10.1103/PhysRevB.82.085114}
}

@article{kamiya2015magnetic,
  author       = {Kamiya, Yoshitomo and Kato, Yasuyuki and Nasu, Joji and Motome, Yukitoshi},
  title        = {Magnetic three states of matter: A quantum {Monte Carlo} study of spin liquids},
  journal      = {Phys. Rev. B},
  volume       = {92},
  pages        = {100403},
  year         = {2015},
  doi          = {10.1103/PhysRevB.92.100403}
}

@article{wu2012phase,
  author       = {Wu, Fengcheng and Deng, Youjin and Prokof'ev, Nikolay},
  title        = {Phase diagram of the toric code model in a parallel magnetic field},
  journal      = {Phys. Rev. B},
  volume       = {85},
  pages        = {195104},
  year         = {2012},
  doi          = {10.1103/PhysRevB.85.195104}
}

@article{reiss2019quantum,
  author       = {Reiss, David A. and Schmidt, Kai Phillip},
  title        = {Quantum robustness and phase transitions of the {3D} toric code in a field},
  journal      = {SciPost Phys.},
  volume       = {6},
  pages        = {078},
  year         = {2019},
  doi          = {10.21468/SciPostPhys.6.6.078}
}

@article{paceWen2023exact,
  author       = {Pace, Salvatore D. and Wen, Xiao-Gang},
  title        = {Exact emergent higher-form symmetries in bosonic lattice models},
  journal      = {Phys. Rev. B},
  volume       = {108},
  pages        = {195147},
  year         = {2023},
  doi          = {10.1103/PhysRevB.108.195147}
}

@article{somoza2021self,
  author       = {Somoza, Andr\'es M. and Serna, Pablo and Nahum, Adam},
  title        = {Self-dual criticality in three-dimensional ${\mathbb{Z}}_{2}$ gauge theory with matter},
  journal      = {Phys. Rev. X},
  volume       = {11},
  pages        = {041008},
  year         = {2021},
  doi          = {10.1103/PhysRevX.11.041008}
}

@article{bonati2022multicritical,
  author       = {Bonati, Claudio and Pelissetto, Andrea and Vicari, Ettore},
  title        = {Multicritical point of the three-dimensional ${\mathbb{Z}}_{2}$ gauge {H}iggs model},
  journal      = {Phys. Rev. B},
  volume       = {105},
  pages        = {165138},
  year         = {2022},
  doi          = {10.1103/PhysRevB.105.165138}
}

@article{stahl2026slow,
  author       = {Stahl, Charles and Placke, Benedikt and Khemani, Vedika and Li, Yaodong},
  title        = {Slow mixing and emergent one-form symmetries in three-dimensional ${\mathbb{Z}}_{2}$ gauge theory},
  journal      = {Phys. Rev. E},
  volume       = {114},
  pages        = {034126},
  doi          = {10.1103/5ypm-x96s},
  year         = {2026},
}

@article{hastings2011topological,
  author       = {Hastings, Matthew B.},
  title        = {Topological order at nonzero temperature},
  journal      = {Phys. Rev. Lett.},
  volume       = {107},
  pages        = {210501},
  year         = {2011},
  doi          = {10.1103/PhysRevLett.107.210501}
}

@article{alicki2009thermalization,
  author       = {Alicki, R. and Fannes, M. and Horodecki, M.},
  title        = {On thermalization in {Kitaev's} {2D} model},
  journal      = {J. Phys. A: Math. Theor.},
  volume       = {42},
  pages        = {065303},
  year         = {2009},
  doi          = {10.1088/1751-8113/42/6/065303}
}

@article{alicki2010thermal,
  author       = {Alicki, R. and Horodecki, M. and Horodecki, P. and Horodecki, R.},
  title        = {On thermal stability of topological qubit in {Kitaev's} {4D} model},
  journal      = {Open Syst. Inf. Dyn.},
  volume       = {17},
  pages        = {1},
  year         = {2010},
  doi          = {10.1142/S1230161210000023}
}

@article{roberts2017symmetry,
  author       = {Roberts, Sam and Yoshida, Beni and Kubica, Aleksander and Bartlett, Stephen D.},
  title        = {Symmetry-protected topological order at nonzero temperature},
  journal      = {Phys. Rev. A},
  volume       = {96},
  pages        = {022306},
  year         = {2017},
  doi          = {10.1103/PhysRevA.96.022306}
}

@article{savaryBalents2017,
  doi          = {10.1088/0034-4885/80/1/016502},
  author       = {Savary, L. and Balents, L.},
  title        = {Quantum spin liquids: a review},
  journal      = {Rep. Prog. Phys.},
  volume       = {80},
  pages        = {016502},
  year         = {2017}
}

@article{nasu2014finite,
  doi          = {10.1103/PhysRevB.89.115125},
  author       = {Nasu, J. and Kaji, T. and Matsuura, K. and Udagawa, M. and Motome, Y.},
  title        = {Finite-temperature phase transition to a quantum spin liquid in a three-dimensional {Kitaev} model on a hyperhoneycomb lattice},
  journal      = {Phys. Rev. B},
  volume       = {89},
  pages        = {115125},
  year         = {2014}
}

@article{nasu2014vaporization,
  doi          = {10.1103/PhysRevLett.113.197205},
  author       = {Nasu, J. and Udagawa, M. and Motome, Y.},
  title        = {Vaporization of {Kitaev} spin liquids},
  journal      = {Phys. Rev. Lett.},
  volume       = {113},
  pages        = {197205},
  year         = {2014}
}

@article{mishchenko2017finite,
  doi          = {10.1103/PhysRevB.96.125124},
  author       = {Mishchenko, P. A. and Kato, Y. and Motome, Y.},
  title        = {Finite-temperature phase transition to a {Kitaev} spin liquid phase on a hyperoctagon lattice: A large-scale quantum {Monte Carlo} study},
  journal      = {Phys. Rev. B},
  volume       = {96},
  pages        = {125124},
  year         = {2017}
}

@article{lu2020detecting,
  doi          = {10.1103/PhysRevLett.125.116801},
  author       = {Lu, Tsung-Cheng and Hsieh, Timothy H. and Grover, Tarun},
  title        = {Detecting topological order at finite temperature using entanglement negativity},
  journal      = {Phys. Rev. Lett.},
  volume       = {125},
  pages        = {116801},
  year         = {2020}
}

@article{wang2025analog,
  author       = {Wang, Ting-Tung and Song, Menghan and Meng, Zi Yang and Grover, Tarun},
  title        = {Analog of topological entanglement entropy for mixed states},
  journal      = {PRX Quantum},
  volume       = {6},
  pages        = {010358},
  year         = {2025},
  doi          = {10.1103/PRXQuantum.6.010358}
}

@misc{chenRouze2025,
  author       = {Chen, Chi-Fang and Rouz\'e, Cambyse},
  title        = {Quantum {Gibbs} states are locally {Markovian}},
  year         = {2025},
  howpublished = {\href{https://arxiv.org/abs/2504.02208}{arXiv:2504.02208}}
}

@article{mcbryanSpencer1977,
  doi          = {10.1007/BF01609854},
  author       = {McBryan, Oliver A. and Spencer, Thomas},
  title        = {On the decay of correlations in {$SO(n)$}-symmetric ferromagnets},
  journal      = {Commun. Math. Phys.},
  volume       = {53},
  pages        = {299},
  year         = {1977}
}

@article{frohlichSpencer1981,
  doi          = {10.1007/BF01208273},
  author       = {Fr\"ohlich, J\"urg and Spencer, Thomas},
  title        = {The {Kosterlitz--Thouless} transition in two-dimensional abelian spin systems and the {Coulomb} gas},
  journal      = {Commun. Math. Phys.},
  volume       = {81},
  pages        = {527},
  year         = {1981}
}

@article{wen1990topological,
  author       = {Wen, Xiao-Gang},
  title        = {Topological orders in rigid states},
  journal      = {Int. J. Mod. Phys. B},
  volume       = {4},
  pages        = {239},
  year         = {1990},
  doi          = {10.1142/S0217979290000139}
}

@article{li2025entanglement,
  author       = {Li, Zhi and Lee, Dongjin and Yoshida, Beni},
  title        = {How Much Entanglement Is Needed for Topological Codes and Mixed States with Anomalous Symmetry?},
  journal      = {Phys. Rev. X},
  volume       = {15},
  pages        = {021090},
  year         = {2025},
  doi          = {10.1103/pw12-kdjx}
}

@article{negari2024spacetime,
  author        = {Negari, Amir-Reza and Ellison, Tyler D. and Hsieh, Timothy H.},
  title         = {Spacetime {Markov} length: a diagnostic for fault tolerance via mixed-state phases},
  journal       = {Phys. Rev. B},
  year          = {2026},
  volume        = {114},
  pages         = {045130},
  doi           = {10.1103/61wf-zx39},
}

@article{zhou2025finite,
  author       = {Zhou, Shu-Tong and Cheng, Meng and Rakovszky, Tibor and von Keyserlingk, Curt and Ellison, Tyler D.},
  title        = {Finite-Temperature Quantum Topological Order in Three Dimensions},
  journal      = {Phys. Rev. Lett.},
  volume       = {135},
  pages        = {040402},
  year         = {2025},
  doi          = {10.1103/n9sq-8cxw},
}

@article{linsel2026paratoric,
  author       = {Linsel, Simon M. and Pollet, Lode},
  title        = {{ParaToric 1.0}: Continuous-time quantum {Monte Carlo} for the toric code in a parallel field},
  journal      = {SciPost Phys. Codebases},
  volume       = {75},
  year         = {2026},
  doi          = {10.21468/SciPostPhysCodeb.75},
}

@book{lahiri2003resampling,
  doi          = {10.1007/978-1-4757-3803-2},
  author       = {Lahiri, Soumendra N.},
  title        = {Resampling Methods for Dependent Data},
  publisher    = {Springer},
  year         = {2003}
}

@article{flammia2009topological,
  author       = {Flammia, Steven T. and Hamma, Alioscia and Hughes, Taylor L. and Wen, Xiao-Gang},
  title        = {Topological entanglement {R\'enyi} entropy and reduced density matrix structure},
  journal      = {Phys. Rev. Lett.},
  volume       = {103},
  pages        = {261601},
  year         = {2009},
  doi          = {10.1103/PhysRevLett.103.261601}
}

@article{dong2008topological,
  author       = {Dong, Shiying and Fradkin, Eduardo and Leigh, Robert G. and Nowling, Sean},
  title        = {Topological entanglement entropy in {Chern--Simons} theories and quantum {Hall} fluids},
  journal      = {J. High Energy Phys.},
  volume       = {2008},
  number       = {05},
  pages        = {016},
  year         = {2008},
  doi          = {10.1088/1126-6708/2008/05/016}
}

@article{affleck1987rigorous,
  author       = {Affleck, Ian and Kennedy, Tom and Lieb, Elliott H. and Tasaki, Hal},
  title        = {Rigorous Results on Valence-Bond Ground States in Antiferromagnets},
  journal      = {Phys. Rev. Lett.},
  volume       = {59},
  pages        = {799--802},
  year         = {1987},
  doi          = {10.1103/PhysRevLett.59.799}
}

@article{kennedy1992hidden,
  author       = {Kennedy, Tom and Tasaki, Hal},
  title        = {Hidden $\mathbb{Z}_2 \times \mathbb{Z}_2$ symmetry breaking in {Haldane}-gap antiferromagnets},
  journal      = {Phys. Rev. B},
  volume       = {45},
  pages        = {304--307},
  year         = {1992},
  doi          = {10.1103/PhysRevB.45.304}
}

@article{thouless1982quantized,
  author       = {Thouless, D. J. and Kohmoto, M. and Nightingale, M. P. and den Nijs, M.},
  title        = {Quantized {Hall} Conductance in a Two-Dimensional Periodic Potential},
  journal      = {Phys. Rev. Lett.},
  volume       = {49},
  pages        = {405--408},
  year         = {1982},
  doi          = {10.1103/PhysRevLett.49.405}
}

@article{haldane1988model,
  author       = {Haldane, F. D. M.},
  title        = {Model for a Quantum {Hall} Effect without {Landau} Levels: Condensed-Matter Realization of the ``Parity Anomaly''},
  journal      = {Phys. Rev. Lett.},
  volume       = {61},
  pages        = {2015--2018},
  year         = {1988},
  doi          = {10.1103/PhysRevLett.61.2015}
}

@article{sandvik2010computational,
  title={Computational Studies of Quantum Spin Systems},
  author={Sandvik, Anders W.},
  journal={AIP Conference Proceedings},
  volume={1297},
  pages={135--338},
  year={2010},
  doi={10.1063/1.3518900}
}

@article{fredenhagenMarcu86,
  author       = {Fredenhagen, K. and Marcu, M.},
  title        = {Confinement criterion for {QCD} with dynamical quarks},
  journal      = {Phys. Rev. Lett.},
  volume       = {56},
  pages        = {223},
  year         = {1986},
  doi          = {10.1103/PhysRevLett.56.223}
}

@article{fredenhagenMarcu83,
  author       = {Fredenhagen, K. and Marcu, M.},
  title        = {Charged states in {$\mathbb{Z}_2$} gauge theories},
  journal      = {Commun. Math. Phys.},
  volume       = {92},
  pages        = {81},
  year         = {1983},
  doi          = {10.1007/BF01206315}
}

@article{gregor2011diagnosing,
  author       = {Gregor, K. and Huse, D. A. and Moessner, R. and Sondhi, S. L.},
  title        = {Diagnosing deconfinement and topological order},
  journal      = {New J. Phys.},
  volume       = {13},
  pages        = {025009},
  year         = {2011},
  doi          = {10.1088/1367-2630/13/2/025009}
}

@article{alles2025FM,
  author        = {All\'es, B. and Borisenko, O. and Chelnokov, V. and Papa, A.},
  title         = {The {Fredenhagen--Marcu} operator in the gauge--{Higgs} {$Z(2)$} lattice gauge theory at finite temperature},
  journal       = {Phys. Lett. B},
  year          = {2026},
  volume        = {880},
  pages         = {140832},
  doi           = {10.1016/j.physletb.2026.140832},
}

@article{chen2010local,
  author       = {Chen, Xie and Gu, Zheng-Cheng and Wen, Xiao-Gang},
  title        = {Local unitary transformation, long-range quantum entanglement, wave function renormalization, and topological order},
  journal      = {Phys. Rev. B},
  volume       = {82},
  pages        = {155138},
  year         = {2010},
  doi          = {10.1103/PhysRevB.82.155138}
}

@article{chen2012symmetry,
  author       = {Chen, Xie and Gu, Zheng-Cheng and Liu, Zheng-Xin and Wen, Xiao-Gang},
  title        = {Symmetry-Protected Topological Orders in Interacting Bosonic Systems},
  journal      = {Science},
  volume       = {338},
  pages        = {1604},
  year         = {2012},
  doi          = {10.1126/science.1227224}
}

@article{coser2019classification,
  author       = {Coser, Andrea and P\'erez-Garc\'ia, David},
  title        = {Classification of phases for mixed states via fast dissipative evolution},
  journal      = {Quantum},
  volume       = {3},
  pages        = {174},
  year         = {2019},
  doi          = {10.22331/q-2019-08-12-174}
}

@article{sang2024mixed,
  author       = {Sang, Shengqi and Zou, Yijian and Hsieh, Timothy H.},
  title        = {Mixed-state quantum phases: Renormalization and quantum error correction},
  journal      = {Phys. Rev. X},
  volume       = {14},
  pages        = {031044},
  year         = {2024},
  doi          = {10.1103/PhysRevX.14.031044}
}

@article{rakovszky2024defining,
  author       = {Rakovszky, Tibor and Gopalakrishnan, Sarang and von Keyserlingk, Curt},
  title        = {Defining stable phases of open quantum systems},
  journal      = {Phys. Rev. X},
  volume       = {14},
  pages        = {041031},
  year         = {2024},
  doi          = {10.1103/PhysRevX.14.041031}
}

@article{sang2025stability,
  author       = {Sang, Shengqi and Hsieh, Timothy H.},
  title        = {Stability of mixed-state quantum phases via finite {Markov} length},
  journal      = {Phys. Rev. Lett.},
  volume       = {134},
  pages        = {070403},
  year         = {2025},
  doi          = {10.1103/PhysRevLett.134.070403}
}

@article{ma2025circuit,
  author        = {Ma, Ruochen and Khemani, Vedika and Sang, Shengqi},
  title         = {Circuit-based characterization of finite-temperature quantum phases and self-correcting quantum memory},
  journal       = {\href{https://arxiv.org/abs/2509.15204}{arXiv:2509.15204}},
  year          = {2025},
}

@article{zhang2025conditional,
  author        = {Zhang, Yifan and Gopalakrishnan, Sarang},
  title         = {Conditional mutual information and information-theoretic phases of decohered {Gibbs} states},
  journal       = {\href{https://arxiv.org/abs/2502.13210}{arXiv:2502.13210}},
  year          = {2025},
}

@article{hastings2007quantum,
  author       = {Hastings, M. B.},
  title        = {Quantum belief propagation: An algorithm for thermal quantum systems},
  journal      = {Phys. Rev. B},
  volume       = {76},
  pages        = {201102},
  year         = {2007},
  doi          = {10.1103/PhysRevB.76.201102}
}

@article{sang2025reversibility,
  author        = {Sang, Shengqi and Lessa, Leonardo A. and Mong, Roger S. K. and Grover, Tarun and Wang, Chong and Hsieh, Timothy H.},
  title         = {Mixed-state phases from local reversibility},
  journal       = {\href{https://arxiv.org/abs/2507.02292}{arXiv:2507.02292}},
  year          = {2025},
}

@article{cong2024enhancing,
  doi          = {10.1038/s41467-024-45584-6},
  pages        = {1527},
  author        = {Cong, I. and Maskara, N. and Tran, M. C. and Pichler, H. and Semeghini, G. and Yelin, S. F. and Choi, S. and Lukin, M. D.},
  title         = {Enhancing detection of topological order by local error correction},
  journal       = {Nat. Commun.},
  volume        = {15},
  year          = {2024}
}

@article{lessa2025strong,
  doi          = {10.1103/PRXQuantum.6.010344},
  author        = {Lessa, Leonardo A. and Ma, Ruochen and Zhang, Jian-Hao and Bi, Zhen and Cheng, Meng and Wang, Chong},
  title         = {Strong-to-weak spontaneous symmetry breaking in mixed quantum states},
  journal       = {PRX Quantum},
  volume        = {6},
  pages         = {010344},
  year          = {2025},
}

@article{zhang2025swssb,
  author        = {Zhang, Carolyn and Xu, Yichen and Zhang, Jian-Hao and Xu, Cenke and Bi, Zhen and Luo, Zhu-Xi},
  title         = {Strong-to-weak spontaneous breaking of $1$-form symmetry and intrinsically mixed topological order},
  journal       = {Phys. Rev. B},
  volume        = {111},
  pages         = {115137},
  year          = {2025},
  doi           = {10.1103/PhysRevB.111.115137}
}

@article{watanabe2026pyrochlore,
  author       = {Watanabe, Sena and Motome, Yukitoshi and Watanabe, Haruki},
  title        = {Dualities and Topological Classification of the {$S=1$} Pyrochlore Spin Ice},
  journal      = {\href{https://arxiv.org/abs/2603.03852}{arXiv:2603.03852}},
  year         = {2026}
}

@article{watanabe2026ice,
  author       = {Watanabe, Sena and Motome, Yukitoshi and Watanabe, Haruki},
  title        = {Continuous crossover between high-pressure ice phases {VII} and {X} driven by monopole screening: A model study},
  journal      = {Phys. Rev. B},
  volume       = {114},
  pages        = {094108},
  year         = {2026},
  doi          = {10.1103/6m5r-8k5m},
}

@article{watanabe2026arbitraryspin,
  author       = {Watanabe, Sena and Motome, Yukitoshi and Watanabe, Haruki},
  title        = {Topological Phase Transitions and Their Thermodynamic Fate in Arbitrary-{$S$} Pyrochlore Spin Ice},
  journal      = {Phys. Rev. X},
  year         = {2026},
  note         = {in press},
  eprint       = {2604.04346},
  archivePrefix = {arXiv},
  primaryClass = {cond-mat.str-el}
}

@article{bao2023mixed,
  doi           = {10.1103/6f98-tvb8},
  author        = {Bao, Yimu and Fan, Ruihua and Vishwanath, Ashvin and Altman, Ehud},
  title         = {Mixed-state topological order and the errorfield double formulation of decoherence-induced transitions},
  journal       = {Phys. Rev. Lett.},
  volume        = {136},
  pages         = {220402},
  year          = {2026},
}

@article{fan2024diagnostics,
  doi           = {10.1103/PRXQuantum.5.020343},
  author        = {Fan, Ruihua and Bao, Yimu and Altman, Ehud and Vishwanath, Ashvin},
  title         = {Diagnostics of mixed-state topological order and breakdown of quantum memory},
  journal       = {PRX Quantum},
  volume        = {5},
  pages         = {020343},
  year          = {2024},
}

@article{jongyeonLee2024coherent,
  doi           = {10.1103/hlfh-86yz},
  author        = {Lee, Jong Yeon},
  title         = {Exact calculations of coherent information for toric codes under decoherence: {I}dentifying the fundamental error threshold},
  journal       = {Phys. Rev. Lett.},
  volume        = {134},
  pages         = {250601},
  year          = {2025},
}

@article{zhao2022measuring,
  author       = {Zhao, Jiarui and Chen, Bin-Bin and Wang, Yan-Cheng and Yan, Zheng and Cheng, Meng and Meng, Zi Yang},
  title        = {Measuring {R\'enyi} entanglement entropy with high efficiency and precision in quantum {M}onte {C}arlo simulations},
  journal      = {npj Quantum Mater.},
  volume       = {7},
  pages        = {69},
  year         = {2022},
  doi          = {10.1038/s41535-022-00476-0}
}

@article{helmes2015renyi,
  author       = {Helmes, Johannes and St\'ephan, Jean-Marie and Trebst, Simon},
  title        = {{R\'enyi} entropy perspective on topological order in classical toric code models},
  journal      = {Phys. Rev. B},
  volume       = {92},
  pages        = {125144},
  year         = {2015},
  doi          = {10.1103/PhysRevB.92.125144}
}

@article{zhao2022scaling,
  author       = {Zhao, Jiarui and Wang, Yan-Cheng and Yan, Zheng and Cheng, Meng and Meng, Zi Yang},
  title        = {Scaling of entanglement entropy at deconfined quantum criticality},
  journal      = {Phys. Rev. Lett.},
  volume       = {128},
  pages        = {010601},
  year         = {2022},
  doi          = {10.1103/PhysRevLett.128.010601}
}

@article{nussinov2008thermal,
  author       = {Nussinov, Zohar and Ortiz, Gerardo},
  title        = {Autocorrelations and thermal fragility of anyonic loops in topologically quantum ordered systems},
  journal      = {Phys. Rev. B},
  volume       = {77},
  pages        = {064302},
  year         = {2008},
}

@article{nussinov2009symmetry,
  author       = {Nussinov, Zohar and Ortiz, Gerardo},
  title        = {Sufficient symmetry conditions for topological quantum order},
  journal      = {Proc. Natl. Acad. Sci. U.S.A.},
  volume       = {106},
  pages        = {16944},
  year         = {2009},
}

@article{xue2024tensor,
  author        = {Xue, Hanyu and Lee, Jong Yeon and Bao, Yimu},
  title         = {Tensor network formulation of symmetry protected topological phases in mixed states},
  journal       = {\href{https://arxiv.org/abs/2403.17069}{arXiv:2403.17069}},
  year          = {2024},
}

@article{lee2023quantum,
  author        = {Lee, Jong Yeon and Jian, Chao-Ming and Xu, Cenke},
  title         = {Quantum Criticality Under Decoherence or Weak Measurement},
  journal       = {PRX Quantum},
  volume        = {4},
  pages         = {030317},
  year          = {2023},
  doi           = {10.1103/PRXQuantum.4.030317},
}

@article{leeYang1952ising,
  author       = {Lee, T. D. and Yang, C. N.},
  title        = {Statistical Theory of Equations of State and Phase Transitions. {II}. Lattice Gas and {Ising} Model},
  journal      = {Phys. Rev.},
  volume       = {87},
  pages        = {410--419},
  year         = {1952},
  doi          = {10.1103/PhysRev.87.410}
}

@article{frohlichRodriguez2017cluster,
  author       = {Fr{\"o}hlich, J{\"u}rg and Rodr{\'i}guez, Pierre-Fran{\c c}ois},
  title        = {On Cluster Properties of Classical Ferromagnets in an External Magnetic Field},
  journal      = {J. Stat. Phys.},
  volume       = {166},
  pages        = {828--840},
  year         = {2017},
  doi          = {10.1007/s10955-016-1556-2}
}

@article{barthel2012quasilocality,
  author = {Barthel, Thomas and Kliesch, Martin},
  title = {Quasilocality and Efficient Simulation of {Markovian} Quantum Dynamics},
  journal = {Phys. Rev. Lett.},
  volume = {108},
  pages = {230504},
  year = {2012},
  doi = {10.1103/PhysRevLett.108.230504}
}

@article{vehtari2021rank,
  author = {Vehtari, Aki and Gelman, Andrew and Simpson, Daniel and Carpenter, Bob and B{\"u}rkner, Paul-Christian},
  title = {Rank-Normalization, Folding, and Localization: An Improved {$\widehat{R}$} for Assessing Convergence of {MCMC} (with Discussion)},
  journal = {Bayesian Analysis},
  volume = {16},
  number = {2},
  pages = {667--718},
  year = {2021},
  doi = {10.1214/20-BA1221}
}

\clearpage
\onecolumngrid
\setcounter{page}{1}
\setcounter{section}{0}
\setcounter{subsection}{0}
\setcounter{subsubsection}{0}
\setcounter{equation}{0}
\setcounter{figure}{0}
\setcounter{table}{0}
\renewcommand{\thesection}{S\arabic{section}}
\renewcommand{\thesubsection}{\thesection.\arabic{subsection}}
\renewcommand{\thesubsubsection}{\thesubsection.\arabic{subsubsection}}
\renewcommand{\theequation}{S\arabic{equation}}
\renewcommand{\thefigure}{S\arabic{figure}}
\renewcommand{\thetable}{S\arabic{table}}
\renewcommand{\thepage}{S\arabic{page}}
\renewcommand{\theHsection}{supp.\arabic{section}}
\renewcommand{\theHsubsection}{\theHsection.\arabic{subsection}}
\renewcommand{\theHsubsubsection}{\theHsubsection.\arabic{subsubsection}}
\renewcommand{\theHequation}{supp.\arabic{equation}}
\renewcommand{\theHfigure}{supp.\arabic{figure}}
\renewcommand{\theHtable}{supp.\arabic{table}}
\makeatletter
\renewcommand{\p@subsection}{}
\renewcommand{\p@subsubsection}{}
\makeatother
\raggedbottom
\pdfbookmark[0]{Supplemental Material}{supplemental-material}
\begin{center}
{\large\bfseries Supplemental Material:\\[0.5ex]
Gibbs-state phases beyond microscopic symmetry breaking:\texorpdfstring{\\}{ }A phase invariant for the 3D toric code}\par
\medskip
Haruki Watanabe
\end{center}
\medskip

This Supplemental Material presents the numerical methods and statistical analysis, followed by the complementary entropy and Fredenhagen--Marcu diagnostics. Equation, figure, and section numbers without an S prefix refer to the main text; notation follows the definitions there.

\section{Quantum Monte Carlo methods and statistical analysis}
\label{sm:numerics}

This section gives the sampling protocols, convergence checks, and uncertainty construction underlying Figs.~\ref{fig:scans}, \ref{fig:scan-moments}, \ref{fig:thermodynamics}, and~\ref{fig:cube}.

Sampling checks test initialization dependence, residual drift, effective sample size, and between-trajectory consistency through split-$\widehat R$. The decoded-correlation cuts pass their prescribed checks, and the survey passes its primary checks, although some auxiliary component checks remain inconclusive. The separate thermodynamic calculations use the protocols below; the transverse-field preparation branches retain points with unresolved chain or block diagnostics.

\subsection{Worldline sampling and the pooled estimator}
\label{sec:fw-implementation}

We use the continuous-time worldline algorithm of Wu, Deng, and Prokof'ev~\cite{wu2012phase}, implemented in ParaToric~\cite{linsel2026paratoric}; Ref.~\cite{sandvik2010computational} reviews the broader quantum Monte Carlo methods for spin systems. For the decoded-correlation calculations, the implementation contains two unglued replicas, each with its own periodic imaginary-time boundary condition. Each retained sample is the complete equal-time $Z$ configuration of one replica.

To help sample different winding parities, the production kernel supplements its worldline updates with global membrane proposals. It chooses one of the $3L$ planar noncontractible dual sheets uniformly, selects either unglued replica with equal probability, and reverses the signs of the $L^2$ link worldlines piercing that sheet throughout imaginary time. This changes the corresponding winding parity while leaving every plaquette product unchanged. The symmetric, involutive proposal is accepted with probability $\min\{1,e^{-\Delta S}\}$, where $\Delta S$ is the change in the integrated diagonal action, preserving the Gibbs sampling weights. At nonzero $h_z$, this acceptance step accounts for the longitudinal-field energy change. These proposals and their global accept/reject decisions are computational sampling steps. The locality restriction applies to the Sweep recovery and to physical preparation channels, not to the Monte Carlo algorithm.

For $N$ retained snapshots, let $w_{i,n}\coloneqq w_i(z_n)$. Hats denote sample estimates, with
\begin{equation}
\widehat M_i\coloneqq\frac{1}{N}\sum_nw_{i,n},\qquad
\widehat D_{12}\coloneqq\frac{1}{N}\sum_nw_{1,n}w_{2,n}.
\label{eq:sample-moments}
\end{equation}
We pool the retained snapshots from all trajectories at the parameter point and form $\widehat f_W\coloneqq\widehat D_{12}-\widehat M_1\widehat M_2$. Since each snapshot contributes to all three moments, uncertainty estimation must resample them jointly. The resampling must also preserve the relevant dependence between samples~\cite{lahiri2003resampling}; the campaign-specific constructions are given below.

\subsection{Sampling units, initialization, and diagnostic checks}
For the decoded-correlation calculations, one Monte Carlo update is one call to the production replica-update kernel, not a whole-lattice sweep or a decoder round. Each call selects uniformly among ten branches: pair insertion/removal and imaginary-time displacement for each of the single-link and star-flip events (four branches); full-imaginary-time reversals of a single link or a star (two); a combined single-link/star-event update; a full-imaginary-time reversal on a noncontractible primal loop; the membrane proposal above; and a flip-event slide across a glued replica boundary. This mixture is used for the five-size cuts, the original survey, and both independent confirmations.

The replica-boundary branch leaves the configuration unchanged because the replicas are unglued. At $h_x=0$, the single-link pair-insertion/removal and displacement branches and the combined single-link/star-event branch likewise leave it unchanged. These calls and rejected proposals count toward the update budget; inactive branches are not removed from the draw. A membrane proposal is therefore selected with probability $1/10$ per call throughout these calculations.

We denote the initial update count by $N_{\rm th}$, the updates between recorded configurations by $N_{\rm sep}$, and the numbers of discarded and retained configurations per trajectory by $N_{\rm drop}$ and $N_{\rm keep}$. The effective discarded evolution is
\begin{equation}
N_{\rm burn}\coloneqq N_{\rm th}+N_{\rm drop}N_{\rm sep}.
\label{eq:burn}
\end{equation}
A trajectory is a Markov chain with a prescribed random-number seed. For these decoded-correlation campaigns, equal numbers of trajectories at each parameter point start from positive and negative winding parities; their seeds are fixed before the run. Balanced starts test convergence to the full Gibbs distribution; they do not prescribe equilibrium sector weights. The moments are pooled over the same retained snapshots before the connected subtraction. Subtracting within each trajectory would omit the covariance of trajectory means. Pooling cannot supply missing equilibrium weights or correct failed equilibration.

Each campaign uses the diagnostic implementations and thresholds specified below. Equivalence requires the interval for a difference to lie within a prescribed tolerance; an unresolved difference does not demonstrate equivalence. Passing these finite-run checks provides evidence of sampling consistency, not a proof of equilibration.

For clarity, the ordinary split-$\widehat R$ is defined as follows. Divide each retained trajectory into its first and second halves and treat these halves as separate sequences. Let $m$ be the number of resulting sequences, each of length $n$, and let $\bar x_j$ and $s_j^2$ be the sample mean and sample variance (with denominator $n-1$) of the monitored scalar series in sequence $j$. With $\bar x\coloneqq m^{-1}\sum_j\bar x_j$, define the within-sequence and between-sequence variances by
\begin{equation}
\begin{aligned}
V_{\rm within}&\coloneqq\frac{1}{m}\sum_{j=1}^{m}s_j^2,\\
V_{\rm between}&\coloneqq\frac{n}{m-1}\sum_{j=1}^{m}(\bar x_j-\bar x)^2.
\end{aligned}
\label{eq:split-rhat-variances}
\end{equation}
For $V_{\rm within}>0$, the diagnostic is
\begin{equation}
\widehat R\coloneqq
\sqrt{\frac{(n-1)V_{\rm within}/n+V_{\rm between}/n}
{V_{\rm within}}}.
\label{eq:split-rhat}
\end{equation}
Splitting makes the comparison sensitive both to differences between runs and to changes within a run. Sampling the same stationary distribution sufficiently well gives $\widehat R$ close to one; an appreciable excess above one warns of incomplete equilibration. It is not a relative error estimate. For the ordinary diagnostic, the implementation assigns $\widehat R=1$ when both variances vanish, whereas $V_{\rm within}=0<V_{\rm between}$ fails the check. The former convention does not establish mixing.

The rank/folded version~\cite{vehtari2021rank} first replaces pooled ranks, with ties assigned their average rank, by standard-normal quantiles and applies the same split-variance calculation. It also repeats this procedure after replacing each sample by its absolute deviation from the pooled median (folding). The larger of the two values is retained to check both location and scale differences. The figure-specific implementations and thresholds are specified below.

Since $f_W$ is a nonlinear combination of three moments, its serial-correlation and between-chain diagnostics use the linearized fluctuation
\begin{equation}
\begin{aligned}
\psi_n\coloneqq{}&w_{1,n}w_{2,n}-\widehat D_{12}\\
&-\widehat M_2(w_{1,n}-\widehat M_1)
-\widehat M_1(w_{2,n}-\widehat M_2).
\end{aligned}
\label{eq:fw-influence}
\end{equation}
All hatted moments here are pooled at the same parameter point. This is the first-order fluctuation of $D_{12}-M_1M_2$ when the three moments vary together. The final estimate and every bootstrap replicate still use the full connected subtraction, not this linearization.

We also check Bianchi closure, total winding parity, and the residual-flux identity in Eq.~\eqref{eq:residual}. These configuration checks validate the recorded and decoded data; they do not alter the correction or discard snapshots with residual flux.


\subsection{Five-size parameter cuts}
\label{sm:fw-cuts}

We use $L=6,8,10,12,14$ on three grids: $T=0.1,0.2,\ldots,2.0$ at $(h_x,h_z)=(0.5,0.1)$; $h_x=0,0.1,\ldots,2.0$ at $(T,h_z)=(0.5,0.1)$; and $h_z=0,0.02,\ldots,0.4$ at $(T,h_x)=(0.5,0.5)$. The three cuts and their size dependence are shown in Fig.~\ref{fig:scans}.

\subsubsection{Sampling budgets}
For each size $L$, let $A$ and $B$ denote the basic update budgets in Table~\ref{tab:fw-cut-budgets}. Ordinary cells use $128$ trajectories with $N_{\rm keep}=1024$, $N_{\rm th}=A$, $N_{\rm sep}=B$, and $N_{\rm drop}=0$.

\begin{table}[htbp]
\caption{Basic update budgets for the five-size cuts. The longitudinal-field cells at $h_z=0.20,0.22,0.24$ use $(N_{\rm th},N_{\rm sep},N_{\rm keep})=(2A,2B,2048)$, except for the independent replacement specified in the text.}
\label{tab:fw-cut-budgets}
\centering
\begin{tabular}{r r r}
\hline
$L$ & $A$ & $B$ \\
\hline
6 & $24\,000\,000$ & $27\,648$ \\
8 & $40\,000\,000$ & $65\,536$ \\
10 & $64\,000\,000$ & $128\,000$ \\
12 & $96\,000\,000$ & $221\,184$ \\
14 & $144\,000\,000$ & $351\,232$ \\
\hline
\end{tabular}
\end{table}

Two cells have additional fixed budgets. At $(L,T,h_x,h_z)=(14,0.70,0.50,0.10)$, a fresh set of $128$ trajectories is pooled with the original $128$, giving $256$ trajectories at the ordinary budget. At $(14,0.50,0.50,0.22)$, a fresh $1536$-trajectory cohort replaces all earlier target data, with $N_{\rm th}=1\,152\,000\,000$, $N_{\rm sep}=702\,464$, and $N_{\rm keep}=2048$. The targets were selected using earlier results, while these follow-up budgets were fixed before their new data were inspected.

\subsubsection{Diagnostic criteria}

The three cuts, including the pooled temperature cell and the independent longitudinal-field replacement, monitor $D_{12},M_1,M_2$, and $f_W$ with the following common thresholds. The absolute first-to-second-half change in each estimate must be at most $0.05$, both after pooling all trajectories and within each initialization separately. These are point-estimate drift checks. For the comparison between initializations, we use only the second halves: the positive-minus-negative difference, plus or minus $1.96$ times its standard error, must lie wholly inside $[-0.05,0.05]$. Its standard error combines, in quadrature, independent delete-one-trajectory jackknife errors for the two initialization groups. Ordinary, untransformed split-$\widehat R$ must be at most $1.05$ for all four monitored quantities; values above $1.01$ are recorded as warnings. The summed trajectory effective sample size must be at least $400$ for each quantity, using $\psi_n$ from Eq.~\eqref{eq:fw-influence} for $f_W$.

The effective-sample-size estimators truncate adjacent-lag autocorrelation pairs at the first nonpositive pair. The retained original $T$ and $h_x$ diagnostics use pairs starting at lag one; the pooled $T$ cell and all $h_z$ diagnostics use pairs starting at lag zero. The corresponding integrated autocorrelation times are bounded below by one, so no trajectory contributes more effective samples than its retained length. A constant individual trajectory contributes zero unless the monitored series is constant across the entire cohort. That latter case is marked constant-resolved by convention and does not provide evidence of mixing. At $h_z=0$, the comparisons between initializations and the split-$\widehat R$/effective-sample-size checks align the signs of the negative-start $M_1,M_2$ series; neither $D_{12}$ nor $f_W$ is sign-flipped, and the pooled physical estimates always use the original values of $w_1,w_2$. All $310$ displayed entries pass these prescribed checks.

\begin{table*}[t]
\setlength{\tabcolsep}{10pt}
\caption{Per-cell budgets for the $L=8$ survey. Counts other than the number of trajectories are per trajectory. Confirmation A targets eight cells, of which five are retained; confirmation B independently replaces the remaining three.}
\label{tab:fw-cube-budgets}
\centering
\begin{tabular}{l r r r r r}
\hline
Calculation & Trajectories & $N_{\rm th}/10^6$ & $N_{\rm sep}$ & $N_{\rm drop}$ & $N_{\rm keep}$ \\
\hline
Stage 1 & 64 & 80 & $65\,536$ & 0 & 2048 \\
Stage 2 & 128 & 320 & $131\,072$ & 0 & 4096 \\
Stage 3 & 128 & 640 & $262\,144$ & 0 & 8192 \\
Confirmation A & 128 & 1920 & $262\,144$ & $16\,384$ & $16\,384$ \\
Confirmation B & 128 & 1920 & $262\,144$ & $131\,072$ & $65\,536$ \\
\hline
\end{tabular}
\end{table*}

\subsubsection{Joint uncertainty and simultaneous intervals}
\label{sm:cut-intervals}

The $310$ panel entries represent $305$ retained data cohorts and $300$ distinct physical conditions: the temperature and transverse-field panels share five reference cohorts, whereas the longitudinal-field panel has separate cohorts there. Uncertainties resample whole trajectories and recompute $D_{12},M_1,M_2$ and their connected subtraction jointly, preserving their covariance, time correlations within each trajectory, and initialization strata. Shared seeds across retained longitudinal-field cells are resampled together; the independent replacement is resampled separately.

For every cell, $5000$ stratified bootstrap replicates give the pointwise standard error of $\widehat f_W$. This standard error is held fixed as the studentizer in a further $10\,000$ max-$T$ replicates for each panel. In each such replicate, we take the largest absolute bootstrap deviation from the observed estimate, divided by its cell's fixed standard error, over the $100$, $105$, or $105$ entries of the respective panel. A zero-standard-error cell contributes zero to this maximum only if every max-$T$ replicate reproduces its point estimate; otherwise the panel intervals remain unresolved. The $95$th percentile of these maxima sets a common critical value, which multiplies each cell's standard error to obtain the displayed symmetric interval. This procedure also applies to the pooled temperature cell and independent longitudinal-field replacement, with the complete respective panel recalibrated.

The bars in Fig.~\ref{fig:scans} are nominal panelwise simultaneous $95\%$ bootstrap intervals: their coverage is approximate, concerns the plotted finite family, and is not guaranteed jointly over all three panels.

For the component moments $D_{12}$ and $M_1$ in Fig.~\ref{fig:scan-moments}, the bars instead use the $2.5$th and $97.5$th percentiles of $5000$ whole-trajectory bootstrap replicates at each cell, stratified by the initial winding sign. These are nominal pointwise $95\%$ intervals for each moment at each parameter point, without the max-$T$ calibration used for Fig.~\ref{fig:scans}.

These intervals quantify sampled fluctuations. Neither construction establishes coverage adjusted for the outcome-informed choice of follow-up targets. Moreover, whole-trajectory resampling cannot detect fluctuations absent from every retained trajectory. Four strongly polarized entries in the $h_z$ panel of Fig.~\ref{fig:scans} have at least one of $w_1,w_2$ exactly constant and hence zero empirical covariance and zero-width bootstrap intervals. Such intervals describe the retained sample; they neither exclude unseen rare changes of $w_1,w_2$ nor establish an exact Gibbs-state value $f_W=0$. The same sampling limitations apply to the component moments and to the survey below. The intervals do not bound systematic bias from incomplete equilibration.


\subsection{Thermodynamic response calculations}
\label{sm:thermodynamic-scans}

Figure~\ref{fig:thermodynamics} combines separate thermodynamic calculations along the same parameter cuts. Let $N_{\rm link}\coloneqq3L^3$, distinct from the sample count $N$ above. For a periodic worldline $z_\ell(\tau)$, define
\begin{equation}
I_z\coloneqq\int_0^\beta d\tau\sum_\ell z_\ell(\tau),\qquad
I_B\coloneqq\int_0^\beta d\tau\sum_p\prod_{\ell\in\partial p}z_\ell(\tau).
\end{equation}
Let $n_x$ and $n_A$ count single-link $X$ events and star-flip events in that worldline. With the stabilizer couplings set to one, the total-energy path estimator is
\begin{equation}
\mathcal E\coloneqq-\frac{h_zI_z+I_B+n_x+n_A}{\beta}.
\end{equation}
Writing $\mathbb E$ and $\operatorname{Var}$ for averages and variances over the equilibrium worldline distribution, the per-link quantities are
\begin{equation}
C_v=\frac{\beta^2\operatorname{Var}(\mathcal E)-\mathbb E(n_x+n_A)}{N_{\rm link}},\qquad
\chi_{zz}=\frac{\operatorname{Var}(I_z)}{\beta N_{\rm link}},\qquad
m_x=\frac{\mathbb E n_x}{\beta h_xN_{\rm link}}.
\label{eq:thermodynamic-estimators}
\end{equation}
The last formula is used at $h_x>0$. The event-count subtraction follows from differentiating the worldline weight at fixed scaled imaginary times. Thus the specific heat is not the variance of an instantaneous diagonal-energy estimator. Similarly, $\chi_{zz}$ is the imaginary-time-integrated static response, not an equal-time spin variance. Numerical estimates replace these moments by pooled sample moments; in panel (b), pooling is restricted to trajectories with the same preparation.

Panels (a,c) use $L=6,8,10,12,14$ along the cuts $(h_x,h_z)=(0.5,0.1)$ and $(T,h_x)=(0.5,0.5)$, respectively. Their parameter grids differ from those of the decoded-correlation scans: each consists of $25$ equally spaced points, including both endpoints of the size-dependent interval in Table~\ref{tab:thermodynamic-grids}, plus two additional points outside that interval. This gives $27$ points per size and cut, or $270$ points in total.

\begin{table}[htbp]
\caption{Thermodynamic sampling intervals for Fig.~\ref{fig:thermodynamics}(a,c). Each interval contains $25$ equally spaced points including both endpoints. Every size additionally includes $T=1.15,1.45$ in panel (a) and $h_z=0.195,0.250$ in panel (c).}
\label{tab:thermodynamic-grids}
\centering
\begin{tabular}{r c c}
\hline
$L$ & $T$ interval & $h_z$ interval \\
\hline
6 & $[1.2500,1.4000]$ & $[0.2100,0.2460]$ \\
8 & $[1.2400,1.3600]$ & $[0.2120,0.2360]$ \\
10 & $[1.2350,1.3250]$ & $[0.2110,0.2320]$ \\
12 & $[1.2350,1.3100]$ & $[0.2130,0.2280]$ \\
14 & $[1.2375,1.2975]$ & $[0.2130,0.2250]$ \\
\hline
\end{tabular}
\end{table}

Each point has eight trajectories, four from each initial winding sign, with $N_{\rm keep}=131\,072$ retained records per trajectory. Initial evolution uses $196\,608N_{\rm link}$ update steps, with $16N_{\rm link}$ steps between records. Each trajectory is divided into $32$ contiguous blocks of $4096$ records; deleting one of the $256$ blocks and recomputing the moments gives the provisional jackknife standard errors.

Diagnostics monitor eight fluctuation series associated with the energy per link, $m_x$, $m_z$, $C_v$, $\chi_{xx}$, $\chi_{zz}$, and the counts per link $n_x/N_{\rm link},n_A/N_{\rm link}$. Here $\chi_{xx}\coloneqq\left.\partial m_x/\partial h_x\right|_{T,h_z}=[\operatorname{Var}(n_x)-\mathbb E n_x]/(\beta h_x^2N_{\rm link})$. For the nonlinear responses $C_v,\chi_{xx},\chi_{zz}$, we use their linearized plug-in fluctuations, obtained by varying their moment formulas to first order about the pooled estimates. These series assess sampling consistency and block adequacy; uncertainties still come from recomputing the full estimators in each jackknife replicate.

All $270$ points pass the prescribed checks. For every monitored series in each trajectory, the integrated autocorrelation time $\tau_{\rm int}$ is measured in records using initial-positive-pair truncation and bounded below by $1/2$. In this convention, the effective sample size is $N_{\rm keep}/(2\tau_{\rm int})$ and must be at least $20$; we also require $2\tau_{\rm int}/N_{\rm keep}\leq0.05$ and $4096\geq10\tau_{\rm int}$. First-to-last-quarter drift is checked separately in every trajectory and must not exceed five standard errors of the difference, using each quarter's variance and autocorrelation-corrected effective sample size. Ordinary split-$\widehat R$ compares the trajectories and must not exceed $1.05$ for each series.

These checks provide finite-run evidence, not an equilibration guarantee. Insets retain maxima over the sampled grids without fitting or interpolation. Their bars are the errors of the selected points; they exclude uncertainty from peak selection and grid resolution.

Panel (b) uses separate single-worldline two-preparation calculations for $L=6,8,10,12,14$ at $(T,h_z)=(0.5,0.1)$. The kernel selects uniformly among the seven single-link, star, and combined single-link/star updates listed above. Each size uses $25$ target fields $h_x=0.76,0.78,\ldots,1.24$ and preparation fields $h_x=0.4$ and $2.2$. Each trajectory starts with all $Z=+1$, evolves for $60\,000$ sweeps at the preparation field, and follows a linear ramp to the target in ten steps of $6000$ sweeps each, including the target as the last step. Here one sweep contains $N_{\rm link}$ attempted Metropolis updates, including rejected proposals. Further evolution at the target precedes measurement; records are then taken every six sweeps.

The displayed data combine three cohorts with different sampling budgets. Of the original $150$ branch points at $L=6,8,10$, $94$ retain four seeds, $60\,000$ target-field settling sweeps, and $48\,000$ records per seed. Both preparations at $28$ size--field pairs were independently recomputed with eight seeds, $600\,000$ settling sweeps, and $192\,000$ records per seed. These pairs comprise the $26$ pairs with unresolved original all-observable diagnostics, plus $L=8,10$ at $h_x=1$. All $56$ remeasured branch points replace their original estimates irrespective of the new diagnostic outcome. The $100$ branch points at $L=12,14$ use the same larger budget. The resulting composite contains $250$ branch points, $1624$ trajectories, and $257\,664\,000$ records, excluding superseded samples. Each plotted point is the equal-weight mean of the seed means for one preparation. All records of the retained trajectories are used; different preparations and original/replacement samples are not pooled.

The panel-(b) error prescription combines between-seed variation with within-seed time correlations. For one branch point, let $n_{\rm seed}\in\{4,8\}$ be the number of seeds and $N_{\rm keep}\in\{48\,000,192\,000\}$ the number of records per seed, as specified above. Let $\bar m_q$ be the mean for seed $q=1,\ldots,n_{\rm seed}$, and let $v_q(b)$ be the sample variance of its $K_b\coloneqq N_{\rm keep}/b$ contiguous block means at block length $b$. Define
\[
\operatorname{SE}_{\rm seed}\coloneqq\frac{\operatorname{sd}(\bar m_1,\ldots,\bar m_{n_{\rm seed}})}{\sqrt{n_{\rm seed}}},\qquad
\operatorname{SE}_{\rm block}(b)\coloneqq\frac1{n_{\rm seed}}\sqrt{\sum_{q=1}^{n_{\rm seed}}\frac{v_q(b)}{K_b}},
\]
where $\operatorname{sd}$ is the sample standard deviation. The displayed provisional standard error is $\max\{\operatorname{SE}_{\rm seed},\operatorname{SE}_{\rm block}(2000)\}$. This prescription is not a confidence interval and does not account for unknown relative equilibrium weights of the two branches.

For each preparation separately, the $m_x$ diagnostics require ordinary split-$\widehat R\leq1.01$ across the $n_{\rm seed}$ seeds. Every seed must have effective sample size at least $100$, $N_{\rm keep}\geq50\tau_{\rm int}$, and $2000\geq10\tau_{\rm int}$. For this scan, $\tau_{\rm int}$ uses initial-positive, monotonized autocorrelation pairs; positive estimates below $1/2$ are retained, with effective sample size $N_{\rm keep}/(2\tau_{\rm int})$. Constant series or nonpositive or undefined estimates remain unresolved. We also require the largest-to-smallest ratio of $\operatorname{SE}_{\rm block}(b)$ at $b=1000,2000,4000$ to be at most $1.25$; an undefined ratio remains unresolved. The diagnostic thresholds are unchanged across cohorts and were fixed before each analysis. Of the $250$ plotted branch points, $248$ pass these $m_x$ checks. The two unresolved points, at $(L,h_x,h_x^{\rm prep})=(6,0.98,2.2)$ and $(8,1.04,0.4)$, are shown as open symbols; here $h_x^{\rm prep}$ denotes the preparation field. All $250$ branch points are retained. Filled symbols indicate only consistency of these within-preparation $m_x$ diagnostics, not global equilibrium convergence. The separation of the two preparations is therefore reported as finite-time preparation dependence, without inferring an equilibrium jump, coexistence field, or transition order.


\subsection{Three-dimensional survey and independent confirmations}
\label{sm:fw-cube}

The $L=8$ survey covers $h_x=0,0.1,\ldots,2.0$, $h_z=0,0.02,\ldots,0.4$, and $T=0.2,0.3,\ldots,2.0$, totaling $8379$ points. The resulting map is shown in Fig.~\ref{fig:cube}.

\subsubsection{Staged sampling and convergence checks}
Table~\ref{tab:fw-cube-budgets} lists the three original stages and the subsequent independent confirmations. A later stage replaces the earlier cohort without pooling. Failed sampling or precision checks, or component drift conclusively outside the equivalence margin, trigger escalation. The map's display threshold does not determine escalation.

The primary $f_W$ checks require rank/folded split-$\widehat R<1.01$, bulk and tail effective sample sizes at least $400$, Monte Carlo standard error at most $0.0125$, and equivalence checks with margin $0.05$. The Monte Carlo standard error is obtained by a delete-one-trajectory jackknife stratified by initialization. For this survey, the rank/folded diagnostics and bulk and tail effective sample sizes~\cite{vehtari2021rank} monitor the linearized fluctuations $\psi_n$ in Eq.~\eqref{eq:fw-influence}. Bulk effective sample size uses the rank-normalized series; tail effective sample size is the smaller value for indicators at the $5$th and $95$th percentiles.

The primary equivalence checks compare the first and second retained halves separately within each initialization group and compare the two initialization groups using their second halves. Differences use delete-one-trajectory jackknife standard errors, retaining the pairing of the two halves from each trajectory and combining independent initialization-group variances. Equivalence requires the two-sided normal interval of nominal level $29/30\simeq96.67\%$ to lie strictly within $(-0.05,0.05)$. This level follows from allocating $0.05/3$ to each of the two one-sided tests at each original look and is retained conservatively in both independent confirmations. These equivalence intervals test sampling consistency. The auxiliary component-equivalence checks apply the same interval construction to $D_{12},M_1,M_2$, using these comparisons and an additional first-to-second-half comparison after pooling the initialization groups. At $h_z=0$, the between-initialization comparison and rank and effective sample size diagnostics for $M_1,M_2$ align the negative-start signs; the pooled physical estimates retain the original values of $w_1,w_2$.

An exactly constant $\psi_n$ series requires an additional check after validating the data and decoder identities. It is admitted only if either $\widehat f_W=0$ with at least one of $w_1,w_2$ constant throughout the retained cohort, or $\widehat f_W=1$ with $w_1=w_2$ at every retained draw and exactly equal numbers of positive and negative common loop values after pooling trajectories separately within each initialization group and each retained half. This convention does not bypass the other checks and does not establish equilibrium or mixing. The equivalence and constant-series rules apply to all original stages and both independent confirmations.

The final composite retains $8282$, $83$, and $6$ original stage-1, -2, and -3 estimates and replaces eight cells by independent confirmations. The confirmation targets were selected using earlier results; each confirmation uses one fixed new cohort, without pooling earlier data. All selected cohorts pass the primary checks, including the constant-series rule above. Auxiliary component-equivalence checks remain inconclusive in $31$ retained original cells; passing the primary checks does not establish equilibration of all observables.

The selected-point views display $\widehat f_W>0.5$ without boundary interpolation. Negative estimates are retained and mapped to the color scale's lower endpoint. This thresholded finite-size map is not a thermodynamic phase boundary.


\section{Quantum and diagonal entropy diagnostics}\label{sm:tee}

We explored two entropy diagnostics: the full quantum R\'enyi-2 entropy and the R\'enyi-2 entropy of equal-time $Z$-basis snapshots. Their distinction matters because the equilibrium snapshot distribution determines diagonal expectation values, whereas subsystem purity also depends on off-diagonal matrix elements.

For a link region $A$, let $\rho_A\coloneqq\Tr_{A^c}\rho$. Its quantum R\'enyi-2 entropy and the entropy of its $Z$-basis distribution are
\begin{equation}
\begin{aligned}
S_2^{\rm q}(A)&\coloneqq-\log\Tr\rho_A^2,\\
S_2^z(A)&\coloneqq-\log\sum_s\langle s|\rho_A|s\rangle^2.
\end{aligned}
\label{eq:tee-entropies}
\end{equation}
Here $A^c$ is the complementary link region, $s$ labels complete $Z$-basis configurations on $A$, and logarithms are natural. The purity obeys
\begin{equation}
\Tr\rho_A^2=\sum_s\langle s|\rho_A|s\rangle^2+
\sum_{s\ne t}|\langle s|\rho_A|t\rangle|^2.
\label{eq:tee-purity}
\end{equation}
Thus $S_2^z\geq S_2^{\rm q}$, with equality precisely when $\rho_A$ is diagonal. This inequality does not imply an ordering of signed entropy combinations. Quantum subsystem entropies are unchanged by products of on-site unitaries, including the path in Sec.~\ref{sec:gibbs-example}; diagonal entropies generally depend on the measurement basis.

Interpreting an entropy combination as topological entropy requires controlled cancellation of local contributions and a large-region limit~\cite{kitaev2006topological,levin2006detecting,castelnovo2008topological}. The field-free value in Eq.~\eqref{eq:field-free-tee} is defined using von Neumann entropies and does not by itself determine a finite-region R\'enyi-2 combination. Region topology matters in three-dimensional gapped ground states~\cite{grover2011entanglement} and Chern--Simons theory~\cite{dong2008topological}, and the temperature and size limits need not commute even in solvable toric codes~\cite{castelnovo2007entanglement}.

Ground-state results require care when applied to these thermal quantities. Topological R\'enyi results for ground states~\cite{flammia2009topological} do not automatically include thermal replica-connectivity contributions such as those found in classical toric codes~\cite{helmes2015renyi}. Spurious contributions can occur even for gapped ground states~\cite{zou2016spurious,williamson2019spurious}. Lower bounds on von Neumann TEE under the assumptions of Refs.~\cite{kim2023universal,levin2024physical} do not imply quantization of the finite-temperature R\'enyi-2 combination.

There is also a distinction between entropy and quantum entanglement in a mixed state. At positive temperature, subsystem entropy includes thermal contributions and is not by itself an entanglement measure. Entanglement negativity~\cite{lu2020detecting} and convex-roof extensions of conditional mutual information~\cite{wang2025analog} address mixed-state entanglement and provide complementary diagnostics.

The full quantum purity can be measured using two replicas~\cite{hastings2010measuring}:
\begin{equation}
\Tr\rho_A^2=\Tr\bigl[(\rho\otimes\rho)\,\mathrm{Swap}_A\bigr],
\label{eq:replica-swap}
\end{equation}
where $\mathrm{Swap}_A$ exchanges the spins in $A$ between the replicas and the trace on the right is over both copies. This exchange changes their imaginary-time boundary connections and includes off-diagonal matrix elements~\cite{humeniuk2012quantum}. Replica quantum Monte Carlo has been applied to mutual information and topological entropy~\cite{melko2010finite,isakov2011topological}; related nonequilibrium increment methods improve entropy estimation~\cite{zhao2022measuring,zhao2022scaling}. Unlike a diagonal-entropy estimate, this calculation must sample the changed replica connections. Controlling their sampling and the uncertainty as the regions grow makes the full quantum calculation more demanding.

Our diagonal-entropy calculations did not establish controlled equilibration or the required large-region limit, and they did not bound the omitted off-diagonal contribution. The full quantum replica calculations also yielded no converged finite-field TEE phase diagram. These outcomes leave the entropy diagnostics unresolved. They do not change the interpretation of $f_W$, whose specified decoded operators are diagonal: their snapshot averages give the full quantum Gibbs expectations.

\section{Fredenhagen--Marcu diagnostic}\label{sm:fm}

We also tested the Fredenhagen--Marcu diagnostic~\cite{fredenhagenMarcu83,fredenhagenMarcu86} in its equal-time spin-system formulation~\cite{gregor2011diagnosing}. It compares an open-string expectation with a closed-loop expectation and provides a complementary candidate to the decoded winding correlation. Finite-temperature applications have also been explored in $(2+1)$-dimensional $\mathbb Z_2$ gauge--Higgs theory~\cite{alles2025FM}. For an open path $\Gamma$ forming half of a contractible contour $C$, define
\begin{equation}
\begin{gathered}
H_\Gamma\coloneqq\left\langle\prod_{\ell\in\Gamma}Z_\ell\right\rangle,\qquad
W_C\coloneqq\left\langle\prod_{\ell\in C}Z_\ell\right\rangle,\\
R_{\rm FM}\coloneqq\frac{|H_\Gamma|}{\sqrt{W_C}},\qquad W_C>0.
\end{gathered}
\label{eq:fm-definition}
\end{equation}
Here the expectations are taken in the Gibbs state. At $h_z=0$, an endpoint star commutes with the Hamiltonian and anticommutes with the open string, giving $H_\Gamma=0$ and hence $R_{\rm FM}=0$ wherever the ratio is defined. This symmetry argument does not extend to $h_z>0$.

We computed this ratio along parameter cuts using L-shaped and U-shaped paths, including independent confirmation runs. The L-shaped scans showed a longitudinal-field response, while the temperature and transverse-field scans remained inconclusive because of small denominators and unresolved sampling convergence. The U-shaped calculations likewise did not resolve the denominator reliably. An unresolved denominator was neither replaced by its absolute value nor assigned a zero ratio. These calculations provide no thermodynamic FM phase boundary; their limitations concern the present numerical estimates and do not establish a failure of the diagnostic itself.

\end{document}